\documentclass[article]{elsarticle}

\usepackage{lineno}
\usepackage{hyperref}
\usepackage{amsthm}
\usepackage{amssymb,amsmath}
\usepackage{graphicx}        % standard LaTeX graphics tool
\usepackage{subfig}  
\usepackage{xcolor}

\usepackage{multirow}  
\colorlet{codecolor}{white!30}
\newcommand{\codebox}[1]{%
  \colorbox{codecolor}{\ttfamily \detokenize{#1}}%
}
\modulolinenumbers[5]

\journal{Journal of Non-Newtonian Fluid Mechanics}

\begin{document}

\begin{frontmatter}

%\title{Piston rheometry}
\title{Numerical assessment of the penetroviscometer approach for large, rapid and transient shear deformations}

%\author{Francisco J. Galindo-Rosales}% \fnref{myfootnote}}
%\address{INEGI, Instituto de Ci{\^e}ncia e Inova{\c{c}}{\~a}o em Engenharia Mec{\^a}nica e Engenharia Industrial, Rua Dr. Roberto Frias, 400, 4200-465, Porto, Portugal; frosales@inegi.up.pt}
%\fntext[myfootnote]{Since 2018.}

\author[mymainaddress1,mysecondaryaddress1]{Ahmad Fakhari}
\address[mymainaddress1]{CEFT, Departamento de Engenharia Mec{\^a}nica, Faculdade de Engenharia da Universidade do Porto, Rua Dr. Roberto Frias s/n, 4200-465, Porto, Portugal}
\address[mysecondaryaddress1]{IPC, Departamento de Engenharia de Pol\'{i}meros, Universidade do Minho, Campus de Azur\'{e}m, 4800-058 Guimar{\~a}es, Portugal}

\author[mymainaddress,mysecondaryaddress]{Francisco J. Galindo-Rosales}% \fnref{galindo@fe.up.pt}}
%\address[mymainaddress]{Instituto de Ci{\^e}ncia e Inova{\c{c}}{\~a}o em Engenharia Mec{\^a}nica e Engenharia Industrial (INEGI), Rua Dr. Roberto Frias, 400, 4200-465, Porto, Portugal}
\address[mymainaddress]{CEFT, Departamento de Engenharia Qu{\'i}mica, Faculdade de Engenharia da Universidade do Porto, Rua Dr. Roberto Frias s/n, 4200-465, Porto, Portugal}
\ead[Corresponding author:]{www.fe.up.pt/ceft - email:galindo@fe.up.pt}

\begin{abstract}
The rheological characterisation of complex fluids is mostly performed under simple shear flow in rotational rheometers. Their modern commercial versions are extremely sensitive instruments which are able to provide very accurate measurements of low values of different material functions, such as viscosity, viscoelastic moduli, etc., under certain ideal flow conditions. Nevertheless, they fail in providing reliable data when characterising the response of complex fluids at short time scales, due to artefacts induced by either instrument or fluid inertia. This is crucial in the analysis of the rheological properties of new formulations of shear thickening fluids specifically developed for protective applications, in which the performance is extremely linked to the time dependent structural changes provoked by the sudden impact loads. Thus, the necessity of providing a reliable experimental tool able to impose large, rapid, transient shear deformation for their rheological characterisation in conditions similar to those of the applications becomes evident. This numerical study aims at assessing the potential use of the \textit{penetroviscometer} for the measurement of the transient shear viscosity of complex fluids at short time scales beyond the current limits of commercial rotational rheometers.  
\end{abstract}


\end{frontmatter}

%\linenumbers

\section{Introduction}
\textit{Rheology} is a field of science dedicated to study the deformation and flow of matter, according to the etymological definition of the word coined by Prof. Bingham in 1920 \cite{Barnesycol1993}. \textit{Rheometry} is dedicated to determine experimentally the rheological properties of complex fluids under well defined and simple flow conditions \cite{GalindoRosales2018}. These standard flows allow sharing and comparing rheological data either for quality control of new formulations, getting insights of the internal structure of the material by comparing with data in the literature or validating constitutive equations\cite{Morrison2001}. Simple shear and extensional flows are complementary standard flow conditions typically considered for the characterisation of a complex fluid, once any complex flow can be split up into components of shear and extensional flows\cite{BarnesMaia2010}. The scientific instrument that allow imposing controlled flow conditions and measure the response of the fluid is named as \textit{rheometer}. Simple shear flow conditions, which only has one non-zero component of the strain rate tensor \cite{Ovarlez2012}, can be achieved experimentally with relative ease either by imposing a pressure difference in a close channel/pipe (pressure driven flow) or by imposing a relative velocity between two solid surfaces (drag flow)\cite{Macosko1994}; and for this reason shear rheometry was developed earlier than extensional rheometry \cite{Birdetal1987a,Galindoetal2013}. Pressure driven rheometers are very useful for determining the viscosity of high viscous materials, nevertheless the lack of homogeneity in the deformation is something that prevents their use for the characterisation of time dependent behaviours\cite{Dealy1998}. Moreover, rheometers based on drag flows, and particularly the rotational rheometers, are much more versatile because they allow to impose different flow kinematics while preserving the homogeneous deformation in the fluid sample and, consequently, providing as a result many different materials functions, and that is the reason their dominant presence in any rheology laboratory worldwide. 
  
Rotational rheometers are equipped with a set of geometries, typically plate-plate, cone-plate and concentric cylinders, having each of them several key features that make them ideal for different kind of fluids and different flow conditions. Concentric cylinders are adequate for low shear rates and low viscosity fluids, while cone-plate allows homogeneous shear rate through out all the volume sample, and plate-plate allows to use different gaps and reach very high shear rate values \cite{Mezger2002,Haake_libro}. All of them allow reaching reliable flow viscosity curves under steady shear flow. These steady shear viscosity curves are obtained by shearing the fluid sample at different shear rates until steady state shear stress response is recorded, or vice versa. This is the typical curve, which demonstrates the viscosity as a function of shear rate $\eta=f\left(\dot\gamma\right)$, used for characterising any complex fluid and fitting the corresponding Generalised Newtonian Fluid (GNF) model, e.g. Carreau, Cross, Bingham, etc. \cite{Morrison2001}, that will be later on used in numerical studies. Despite this is a correct methodology for characterising complex fluids and predicting their flow behaviour under steady flow conditions, it may produce misleading results when used for predicting the transient response of these fluids. GNF models assume that there is no time delay between the applied shear rate and the change in the viscosity of the fluid; nevertheless, most of complex fluids from colloidal suspensions to polymer solutions shows memory effects, exhibiting a change in the viscosity that is not synchronised with the application of the shear rate. In other words, they exhibit a viscoelastic response, where the elastic component is represented by the response in phase with the deformation ($G'$) and the viscous component is given by the response in phase with the rate of deformation ($G''$). This is very relevant in the case of shear thickening fluids (STFs), which are typically used in shock absorbing applications where the flow conditions are intrinsically transient \cite{GalindoRosales2015326,FJGRLCDPatent,galindoApplSci2016}. Despite STFs exhibit both viscoelastic moduli ($G'$ and $G''$)\cite{Goede2019}, they are typically characterised under steady shear flow conditions and their viscosity is the key parameter used for designing shock absorbing devices \cite{GURGEN2016312, GURGEN2018, KHODADADI2019643}. It is hard to correlate the mechanical response of the shock absorbers in very short time scales with the rheological information obtained under steady flow conditions, even more when it has been widely reported that STFs cannot instantaneously change their rheological properties from liquid like to solid like due to its viscoelastic nature. The paradigm is currently changing and new studies are published reporting that the characterisation of STFs for improving the mechanical properties of composites should be done in terms of the \textit{instantaneous viscosity} instead of the steady state viscosity \cite{Pinto2017}. However, the problem is to determine how instantaneous is \textit{instantaneous viscosity} measured of a STF in a rotational rheometer.    

In rotational rheometers, there are two major limitations when characterising the mechanical response of complex fluids at short time scales, which are the instrument inertia and fluid inertia. In the seminal book chapter by Ewoldt et al. \cite{Ewoldt2015}, it is clearly stated that instrument inertia limits the maximum frequency at which a rotational rheometers can provide reliable data ($\omega<\sqrt{\frac{GF_{\gamma}}{IF_{\tau}}}$), being $G$ either $G^\prime$ or $G^{\prime\prime}$, and $\frac{IF_{\tau}}{F_{\gamma}}$ the instrument inertia associated to the measurement geometry); they also clearly expose that instrument inertia affects the minimum acquisition time to provide reliable data in step tests due to the instrument acceleration at short-time scale ($t>\sqrt{2\frac{\eta^{+}F_{\tau}I}{F_{\gamma}}}$, where $\eta^{+}$ is the transient shear viscosity\cite{nomenclature}). Additionally, one has to consider the fluid inertia associated to the secondary flows, due to curved streamlines and high velocities; as well as the presence of a wave propagating through the volume sample as a result of either viscous momentum diffusion or elastic shear waves or both, which wavelength $l$ should be much greater than the geometry gap $D$ in order to avoid this artifact ($l\gg D$). Thus, it becomes evident that conventional rheometers have the disadvantage of being too massive, resulting in inertial problems and limiting their range of operation to relatively low frequencies. New approaches would be required to determine the \textit{instantaneous viscosity}.

The sliding plates rheometers, conversely to rotational rheometers, have been reported to be a successful approach for characterising shear thinning fluids under ``large, rapid, transient shear deformations'' \cite{Dealy1998,Giacomin1989,ORTMAN2011884} allowing to reach high frequencies when scaling down in the gap size (sliding microrheometers) \cite{CLASEN20041, Moon2008, Verbaan2015}. Thanks to plane Couette flow conditions, it is also possible to perform simultaneously non-mechanical measurements, i.e. neutron scattering analysis. However, not everything is perfect on slide rheometry, and wall-slip issues may arise\cite{Hatzikiriakos1991}. Despite the slide rheometry looks like a good approach for fast transient measurements, to the best of authors knowledge it has never been used for characterising shear thickening samples at short time scales, probably due to problems related with keeping the gap size constant, overloading issues of the transducer, shear fracture or even wall-slip \cite{Dealy1998}.  

The approach in the sliding cylinders rheometer is similar to the sliding plate rheometer, but it prevents the edge effects and bearing friction issues. The former shares the same principle with the falling rod viscometer \cite{Dealy1998}, which is considered a precise method for measuring the absolute viscosity of Newtonian fluids ranging from $10^{-3}$ to $10^{7}$ Pa$\cdot$s \cite{debruyn}. In both cases, when the relative gap between the cylinders is very small, there is no need to know the constitutive equation of the fluid to calculate the shear strain and shear rate, as in the sliding plates rheometer\cite{Dealy1998}. 

In 1948, Bikerman \cite{BIKERMAN194875} proposed the \textit{penetroviscometer} as a new viscometer for determining the viscosity of Newtonian fluids with a viscosity between $100$ and $100,000$ Pa$\cdot$s under steady state conditions with a remarkable error below 3.5\%. The configuration is similar to the falling rod viscometer in the sense that the fluid is contained between two coaxial cylinder, being the outer one fixed and the inner one movable; but they are different because the area of contact between the liquid and the inner cylinder is not constant in the \textit{penetroviscometer}. Besides, another difference is that in the viscosimeter proposed by Bikerman, when the inner cylinder moves downward, the fluid is forced to flow upward through the annular gap between the two cylinders. Bikerman's analysis was done for measurements on steady state and to the best of authors knowledge it has never been assessed for the measurement of the transient shear viscosity of complex fluids at large, rapid, transient shear deformations. From the experimental point of view, it would be extremely easy to convert a standard drop weight impact machine into a penetroviscometer, just by recording the position of the tip of the inner cylinder with time and measuring the force by means of a piezoelectric force transducer located at the shaft of the inner cylinder. In this way, deformation and the stresses in the fluid will be decoupled, as in a separated motor-transducer instruments, helping in avoiding the instrument inertia effect \cite{Ewoldt2015}.

In this work, we perform a numerical analysis on the usefulness of the penetroviscometer-like viscometer for measuring transient viscosities of complex fluids under  large, rapid, transient shear deformations. Our attention is given to fluids with a viscosity ranging from $10^{-3}$ to $10^{3}$ Pa$\cdot$s and we analyze different potential experimental conditions, such as the ratio between the radius of the concentric cylinders, the velocity of the inner cylinder (in case it was a control parameter experimentally) and the initial position of the inner cylinder's tip.
  
\section{Materials and methods}
 
\subsection{Geometry and initial conditions}

\begin{figure}[t]
%\sidecaption
\centering
% For example, with the graphicx style use

{\label{fig: Sketch}\includegraphics[width=0.75\linewidth]{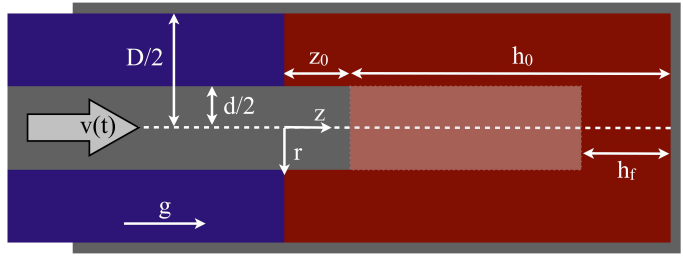}}

\caption{Sketch of the penetroviscometer. The volume of the reservoir filled with blue represents the air, while the red color represents the liquid to be tested. The reference of the coordinate system is located at the interface of the liquid before the impact.}
\label{fig:1}       % Give a unique label
\end{figure}

The geometry consists of a stationary cylindrical reservoir with an inner diameter $D$, and a sliding cylinder (inner cylinder) coaxial with the reservoir and with a smaller diameter ($d$), which moves inside it in the direction of the gravity ($g$). The origin of the coordinate system is located at the axis of the cylinders and at the same height as the interface between the two fluids before the experiment starts. A sketch of the geometry is shown in Fig. \ref{fig:1}. Three different geometries are considered for this study, based on three different blockage ratios (BR=$d/D=1/1.5,~ 1/3$ and $1/6$). 

At time $t=0$, it is considered that the tip of the inner cylinder is submerged in the liquid a distance $z_{0}$, being $z_{0}=0,~ d, ~2d,~ 16d$. Its initial velocity is 1.2 m/s, which is a typical velocity value for drop weight impact machines. Additionally, both fluids are considered to be at rest and the height of the liquid is $h_{0}$.

\subsection{Governing equations}

The governing equations are the mass conservation equation (Eq. \ref{eq: Continuity}):
 
   \begin{equation}
  \label{eq: Continuity}
  \dfrac{\partial \rho}{\partial t} + \nabla \cdot \rho U =0, 
  \end{equation} 
 
\noindent and the momentum conservation equation (Eq. \ref{eq: Momentum}):
 
   \begin{equation}
  \label{eq: Momentum}
  \dfrac{\partial \left( \rho U \right)}{\partial t} + \nabla\cdot \left( \rho UU \right) =-\nabla P + \nabla \cdot  \tau + \rho g + f_s,
  \end{equation}  

\noindent where $U$ is the velocity vector shared between the two fluids in the entire domain, $P$ is the pressure, $\tau=2\mu S - 2 \mu \left( \nabla \cdot U \right) I /3$ is the deviatoric viscous stress tensor, $S=1/2 \left[\nabla U+ \left( \nabla U \right)^T \right]$ is the strain rate tensor, $I$ is the unity matrix. 

The interface liquid-air will be treated by considering the volume of fluid (VOF), which is a powerful method to approximate free boundaries in finite-difference numerical simulations. It was proposed by Hirt and Nichols \citep{HirtN1981} and it uses a scalar function ($\alpha$), called volume fraction, to define if the region is occupied by the liquid ($\alpha=1$), empty of that liquid ($\alpha=0$) or if it corresponds to a free surface ($0<\alpha<1$). Therefore, Eq. \ref{eq: Continuity} and Eq. \ref{eq: Momentum} will be solved in combination with the transport equation for volume fraction (Eq. \ref{eq: MixtureFraction}):
  
  \begin{equation}
  \label{eq: MixtureFraction}
  \dfrac{\partial \alpha}{\partial t} + \nabla \cdot \left( U  \alpha  \right)= 0,
  \end{equation} 
  
\noindent in which $\alpha$ varying in the range of $0 \leq \alpha \leq 1$. The surface tension $f_s$ is also calculated as follows \cite{Andersson2010}:
    
    \begin{equation}
    f_s=\sigma\left( \nabla \cdot \left( \dfrac{\nabla \alpha}{|\nabla \alpha|} \right) \right) \left( \nabla \alpha \right),
    \end{equation}

    %, here $\sigma=0.07$ is used for all the simulations,
\noindent where $\sigma$ is the interface coefficient and $\nabla \alpha=n$ is the vector normal to the interface \cite{Andersson2010}. The fluid properties which are density and  dynamic viscosity can also be obtained at each computational cell using the volume fraction
     
     \begin{equation}
     \rho=\alpha \rho_l+ \left( 1 - \alpha \right) \rho_a,
     \end{equation}
      \begin{equation}
     \mu=\alpha \mu_l+ \left( 1 - \alpha \right) \mu_a,
     \end{equation}
     
\noindent here indexes $l$ and $a$ are representative of liquid and air, respectively. 
 
 \subsection{Boundary conditions}
 
\begin{figure}[h!]
%\sidecaption
\centering
% For example, with the graphicx style use

\includegraphics[width=\linewidth]{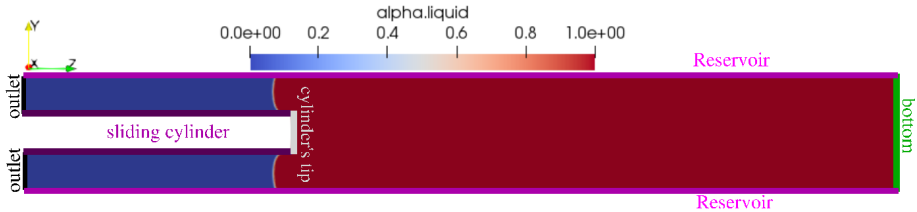}

\caption{A slice of the computational domain and the boundaries which are shown with differentcolors, BR=1/3 $z_0=d$.}
\label{fig:2}       % Give a unique label
\end{figure}

 Fig. \ref{fig:2} displays a slice of the computational domain and the boundaries. No-slip boundary condition with zero velocity at the side and bottom walls of the reservoir is imposed. At the lateral surface and at the tip of the sliding cylinder, again no-slip boundary condition, but the velocity is given by $v(t)$. For the velocity at the outlet \codebox{pressureInletOutletVelocity} is imposed, in which zero-gradient is applied and the velocity is obtained from the patch-face normal component of the internal-cell value \cite{Weller1998}.\\
 
For pressure, \codebox{zeroGradient} boundary condition is applied at all the boundaries except at the outlet, where a total pressure equal to zero is set. For the mesh motion, the velocity $v(t)$ is given to the sliding cylinder and its tip in \codebox{pointMotionUz} file, while no-slip boundary condition (\codebox{fixedValue uniform (0 0 0)}) is set at the reservoir. A uniform fixed value equal to zero is also used at the bottom and outlet.

Three different functions for the velocity have been considered for the sliding cylinder $v(t)$:
 
 \begin{itemize}
 \item Constant: $v(t)=1.2$ m/s;
 \item Linear: $v(t)=1.2-29.09 t$  m/s;
 \item Parabolic: $v(t)=1.2 -1.84\cdot 10^{-4}t -705 t^2$ m/s. 
 \end{itemize} 

\begin{figure}[h]
\centering
\includegraphics[width=0.9\linewidth]{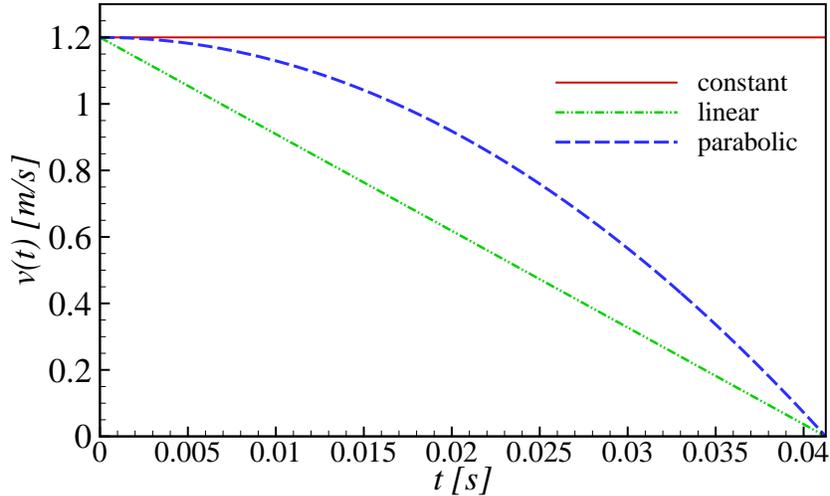}

\caption{Different velocity profiles $v(t)$ considered for the impactor in the assessment of the penetroviscosimeter to measure the transient viscosity of the liquid.}
\label{fig:velocityprofiles}        % Give a unique label
\end{figure}

These velocity profile have been considered based on the results of some preliminary experiments \cite{Rafael2017}, where the initial impact velocity was 1.2 m/s and the time evolution of the impactor's velocity followed a parabolic deceleration until it was stop after $\sim40$ ms. The linear velocity profile was defined considering the same initial velocity and the 40 ms to be stopped (Fig. \ref{fig:velocityprofiles}). 
   \\

    \begin{table}[h!]
\caption{Computational domain length consist of the length of the sliding cylinder at $t=0$ s and the length downstream of the impactor, number of grids in stream-wise and the ratio of the smallest to the largest cell ($R_c$) for the cases with $z_0=0$ and $z_0$ $\neq$ 0.} % title of Table
\centering % used for centering table
\begin{tabular}{c c c} % centered columns (4 columns)
\hline\hline %inserts double horizontal lines
Impactor's velocity & $L_{impactor_{t_0}}$  & $L_{downstream}$ \\ [0.5ex] % inserts table
& [cells, $R_c$]  & [cells, $R_c$] \\ [0.5ex]
\hline % inserts single horizontal line
constant ($z_0=0$) & $10d$[320,0.8] & $16.4d$[80,0.05] \\ % inserting body of the table
linear ($z_0=0$)& $5d$[160,0.8] & $16.4d$[80,0.05] \\
parabolic ($z_0=0$)& $6.7d$[213,0.8] & $16.4d$[80,0.05]\\ [0.5ex] % [1ex] adds vertical space
\hline
parabolic: $z_0=d$ & $7.7d$[120,0.8] & $16.4d$[40,0.05]\\ [0.5ex] % [1ex] adds vertical space
% %inserts single line
parabolic: $z_0=2d$ & $8.7d$[136,0.8] & $16.4d$[40,0.05]\\ [0.5ex] % [1ex] adds vertical space
% %inserts single line
parabolic: $z_0=16d$ & $22.7d$[320,0.8] & $16.4d$[40,0.05]\\ [0.5ex] % [1ex] adds vertical space
\hline %inserts single line
linear (doubled resolution) & 5d[320,0.8] & $16.4d$[160,0.05]\\[0.5ex]
\hline
\end{tabular}
\label{table: domain sizes} % is used to refer this table in the text
\end{table}

\subsection{Numerical considerations}

OpenFOAM 2.4.0 is used to numerically model the fluid flows between the two cylinders. Among the OpenFOAM's solvers,  \codebox{multiphaseInterDyMFoam} is chosen to impose the motion of sliding cylinder (impactor) inside reservoir applying dynamic mesh and solve the Navier-Stokes equations for a multiphase flow. This solver applies volume of fluid (VOF) to capture the interface of the fluids (here air and liquid), as it will be briefly discussed in the following section.

%\textcolor{blue}{COMMENT: We must prove grid independency (Table \ref{table: domain sizes}).}

The \codebox{multiphaseInterDyMFoam} solver was employed to handle dynamic mesh for solving Navier Stokes equations of two phases flow consisting of air ($\mu=1.8375\cdot10^{-6}$ Pa$\cdot$s and $\rho=1.225$ kg/m$^{3}$) and the liquid. The simulations are carried out starting from a very viscous liquid (silicon-1000), continued with another fluid with a less viscosity (silicon-1), and accomplished with water. The key properties of the three different liquids are given in Table \ref{table: liquids}.

    \begin{table}[h!]
\caption{Physical properties of the working fluids used in this study.} % title of Table
\centering % used for centering table
\begin{tabular}{c c c c} % centered columns (4 columns)
\hline\hline %inserts double horizontal lines
\multirow{2}{*}{Fluid}& Dynamic viscosity     & Density  			& Surface tension  \\ [0.5ex] % inserts table
	 & $\mu$ [Pa$\cdot$s]  & ($\rho$ [kg/m$^{3}$]) & ($\sigma$ [mN/m])\\ [0.5ex]
\hline % inserts single horizontal line
Silicon-1000 & 1000 & 970 & 35 \\ % inserting body of the table
Silicon-1 & 0.970 & 970 & 35 \\ % inserting body of the table\hline
Water & 0.000997 & 997 & 70 \\ % inserting body of the table
\hline %inserts single line
\end{tabular}
\label{table: liquids} % is used to refer this table in the text
\end{table}

\begin{figure}[h!]
\centering
\subfloat[Mesh corresponding to the fluid domain at the interface liquid-air.]{\label{fig:outlet}\includegraphics[width=0.4\linewidth]{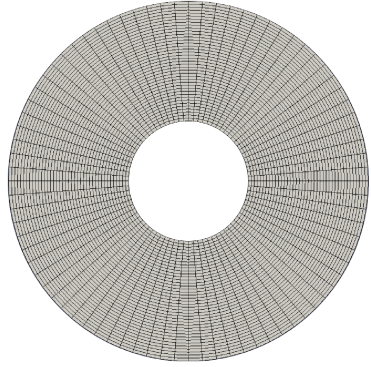}}\hfill
\subfloat[Mesh corresponding to the fluid domain at the bottom of the reservoir.]{\label{fig:bottom}\includegraphics[width=0.4\linewidth]{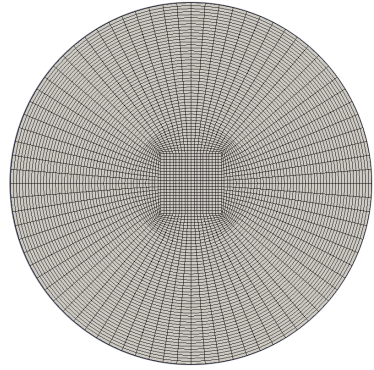}}\\

\subfloat[Mesh corresponding to the sliding cyliner's tip.]{\label{fig:impactorTip}\includegraphics[width=0.25\linewidth]{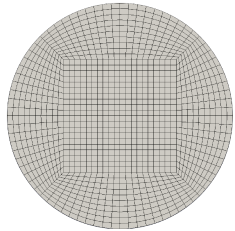}}\hfill
\subfloat[A longitudinal cut view of a the domain including the sliding cylinder.]{\label{fig:cutViewOfTheDomain}
\includegraphics[width=0.6\linewidth]{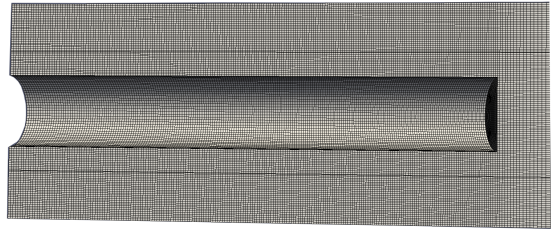}}

\subfloat[Side view of the mesh used for the reservoir.]{\label{fig:cutViewOfTheDomain}
\includegraphics[width=\linewidth]{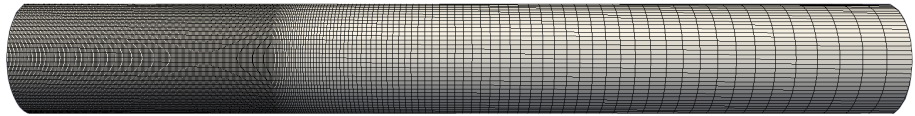}}

\caption{Computational domain and mesh for the case in which BR=1/3 and $z_0=0$.}
\label{fig:numericaldomain}        % Give a unique label
\end{figure}

Cylindrical polar coordinates are considered ($r,z,\theta$). The structured mesh is generated using blockMesh utility. The cross-section of the sliding cylinder is created and meshed as a combination of a $d/2 \times d/2$ square at the center with a resolution of $20 \times 20$ cells, and then each side of the square is transformed to the impactor's edge by 10 for the cases with BR=1/1.5 and $z_0=0$. A coarser grid is used for the cases with $z_0 \neq 0$; the same square size  ($d/2 \times d/2$) covered by $12 \times 12$ cells and the area between the square and the inner cylinder's edge is covered by 6 cells for the case with BR=1/1.5. The grid spacing size is kept the same for the other two blockage ratios. Fig. \ref{fig:numericaldomain} shows some details of the numerical domain and Table \ref{table: domain sizes} provides the domain length, number of cells in stream-wise ($z$) direction, and the ratio of the smallest to the largest cells. 

While the grid is uniform in the radial direction, a constant stretch ratio is applied in the stream-wise direction ($z$) to have highest resolution at the impactor's tip, where there is air/liquid interface for $z_0=0$. The simulations are carried out using a fixed time step $\Delta t=1.375$ $\mu$s, with the simulation time equal to $41$ ms. Euler time integration is applied and the data is printed every $1.375$ ms. Gauss linear discretization scheme is used for the derivatives, except  $\nabla \cdot \left (U \alpha \right)$, for which Van Leer divergence scheme is employed. %The drag force ($F_{w}=\tau_{w}\cdot A_{L}$), due to the friction exerted by the liquid at the lateral wall of the inner cylinder ($\tau_{w}=\eta\cdot\dot\gamma_{w}$), is computed at every time step and compared with the pressure force at the bottom of the inner cylinder ($F_{P}=P\cdot \pi \frac{d^2}{4}$). 

%\textcolor{blue}{COMMENT: Ahmad,I also think that we should add a figure showing all the details of the domain and the mesh Fig.\ref{fig:numericaldomain}.}\textbf{I have added some figures but I am not sure what we should exactly show. I'll send you a zipped folder containing the mesh so that you can also have a look for having a better idea. }

%\subsubsection{Numerical method}

 %The drag force ($F_{w}=\tau_{w}\cdot A_{L}$), due to the friction exerted by the liquid at the lateral wall of the sliding cylinder (the wall shear stress $\tau_{w}=\eta\cdot\dot\gamma_{w}$), is computed at each time step and compared with the pressure force at the sliding cylinder's tip ($F_{P}=P\cdot \pi \frac{d^2}{4}$). 

%\subsubsection{Mesh analysis}
%\textcolor{blue}{This is important to provide credibility to our results. I believe that we need to define a sort of reference parameter and compare the results obtained for that parameter with the different meshes, right? Could you please do this analysis?}
%\textcolor{red}{grid convergence study needs to be completed. We have the simulation with double resolution of the case in which the impactor's velocity is linear ($z_0=0$), but I don't know if this is enough. I still need to plot the forces for this case and compare it with the coarser grid which is the second row in Table \ref{table: domain sizes}.}

    \begin{figure}
%\sidecaption
\centering
% For example, with the graphicx style use

\includegraphics[width=0.75\linewidth]{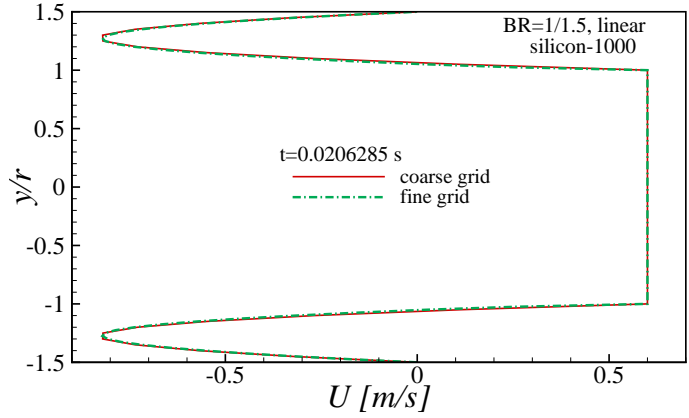}

\caption{Velocity profile near the sliding cylinder's tip for silicon-1000, BR=1/1.5 with linear imposed velocity to the sliding cylinder; a comparison between the coarse and fine (doubled resolution) grids at $t \sim 2$ ms.}
\label{fig: gridAnalysis}       % Give a unique label
\end{figure}

%\subsection{Calculus of the instantaneous viscosity}

\begin{figure}[h!]
\centering 
\subfloat[]{\label{fig: Forces_silicon_BR1.5_z0=0}\includegraphics[width=0.33\linewidth]{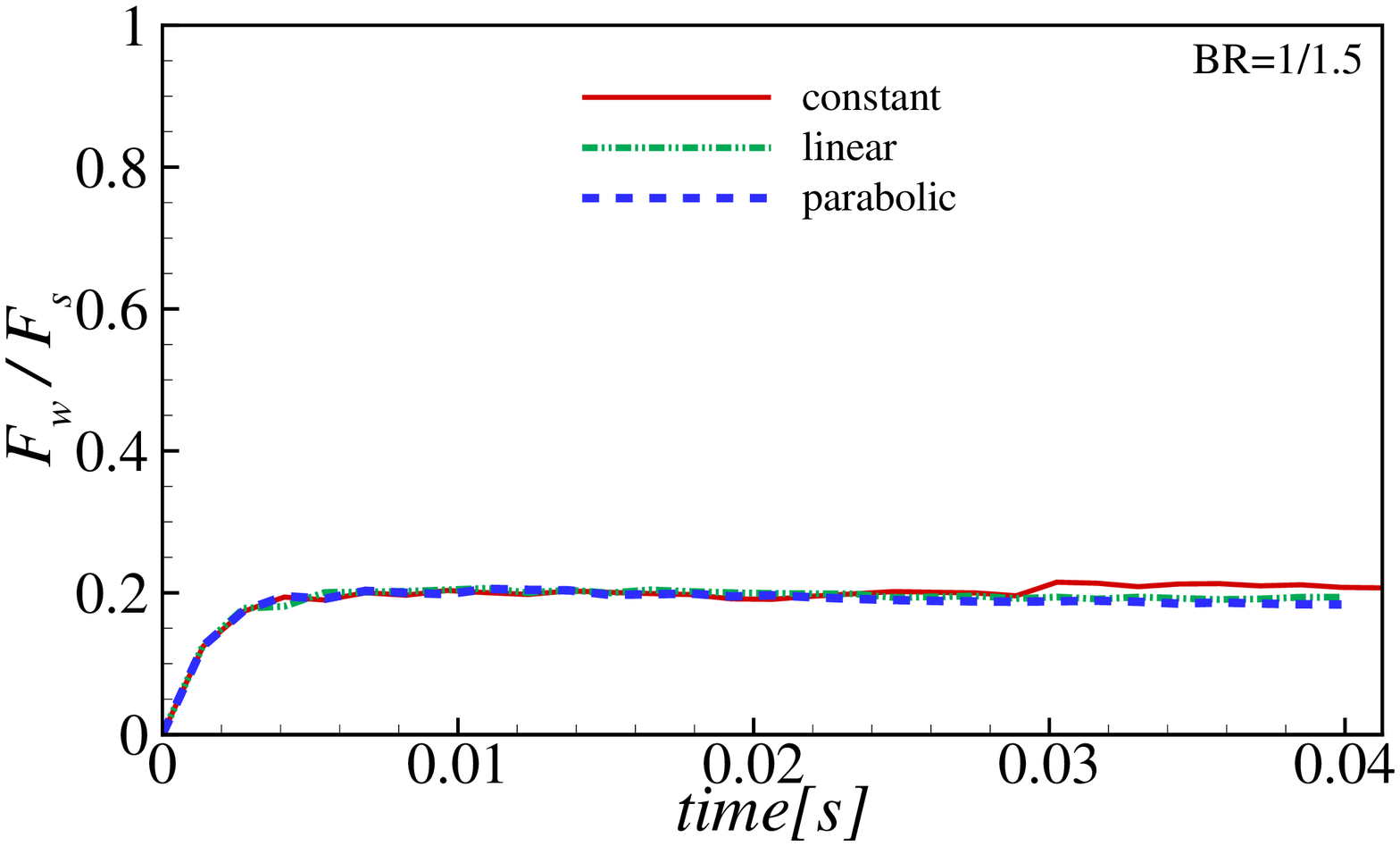}}
\subfloat[]{\label{fig: Forces_silicon_BR3_z0=0}\includegraphics[width=0.33\linewidth]{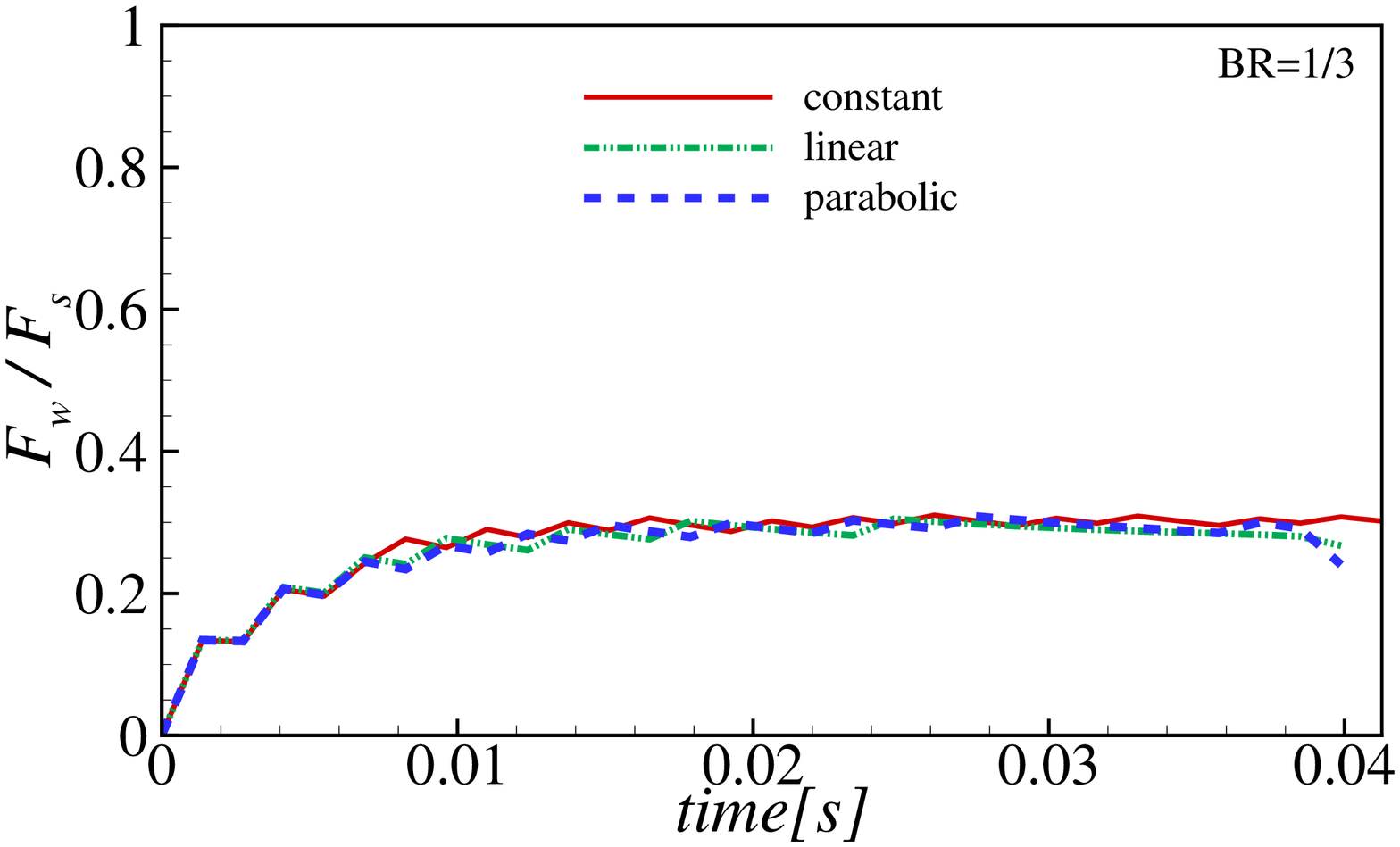}}
\subfloat[]{\label{fig: Forces_silicon_BR6_z0=0}\includegraphics[width=0.33\linewidth]{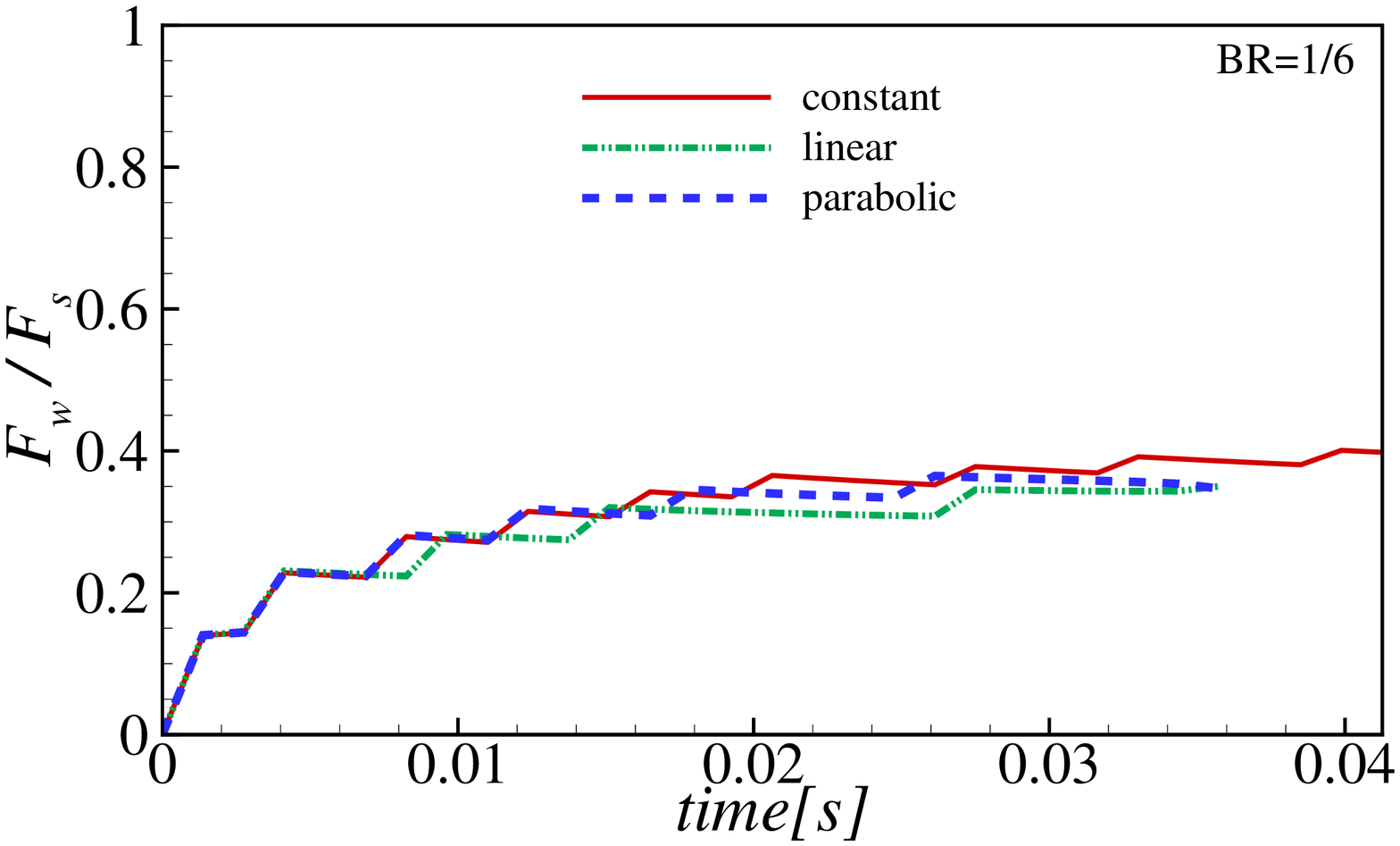}}

\subfloat[]{\label{fig: Forces_silicon_BR1.5_nu0001_z0=0}\includegraphics[width=0.33\linewidth]{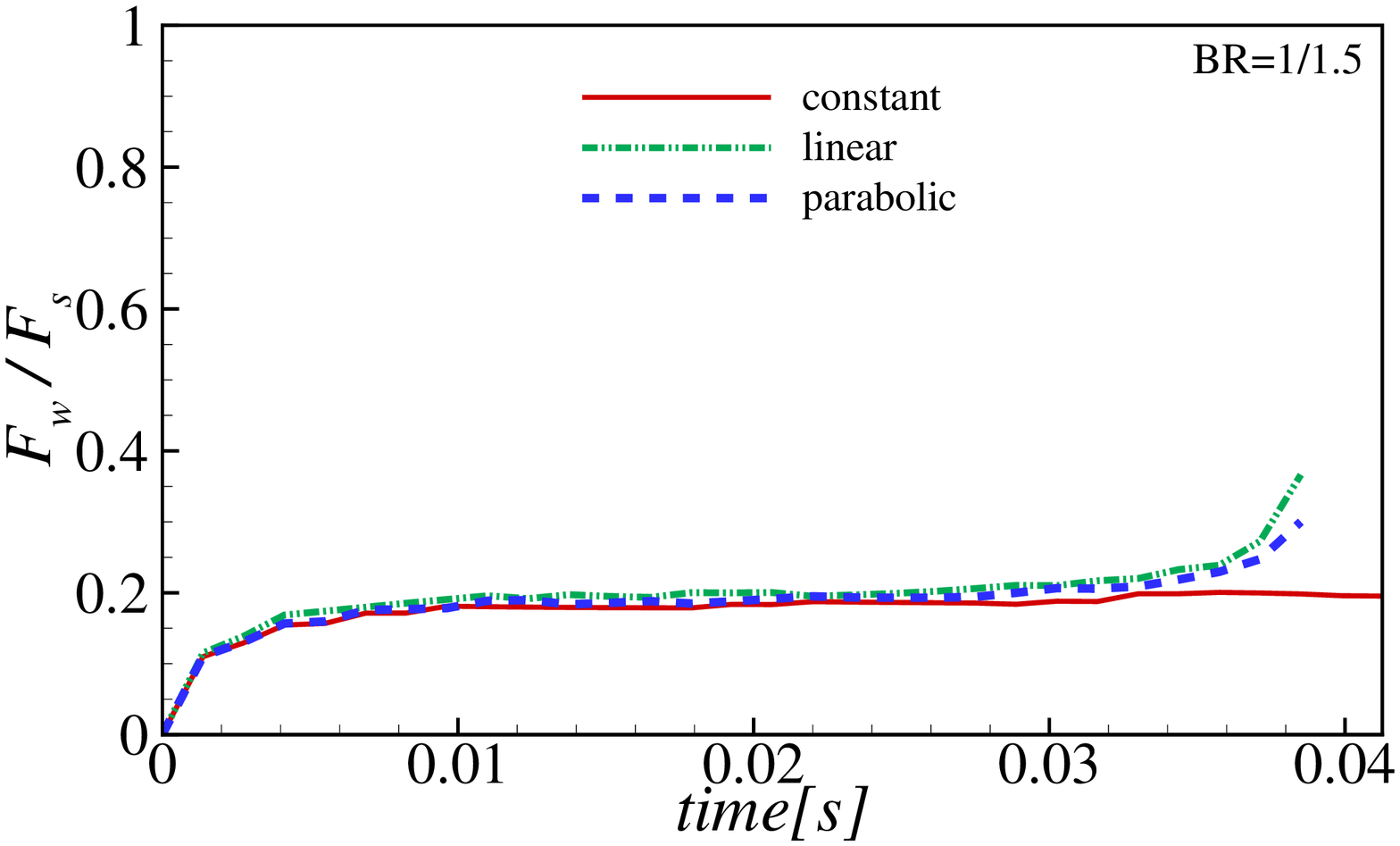}}
\subfloat[]{\label{fig: Forces_silicon_BR3_nu0001_z0=0}\includegraphics[width=0.33\linewidth]{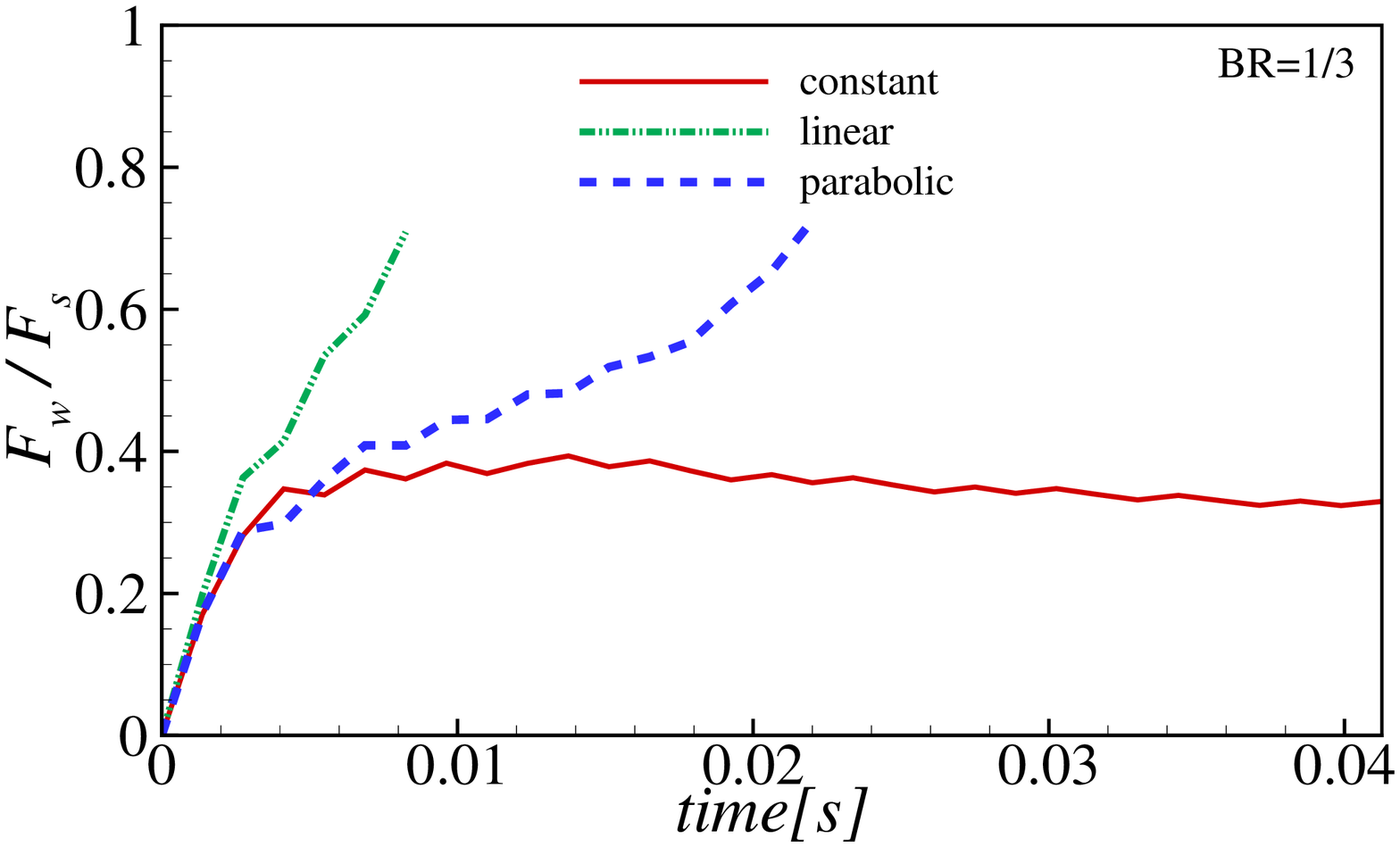}}
\subfloat[]{\label{fig: Forces_silicon_BR6__nu0001_z0=0}\includegraphics[width=0.33\linewidth]{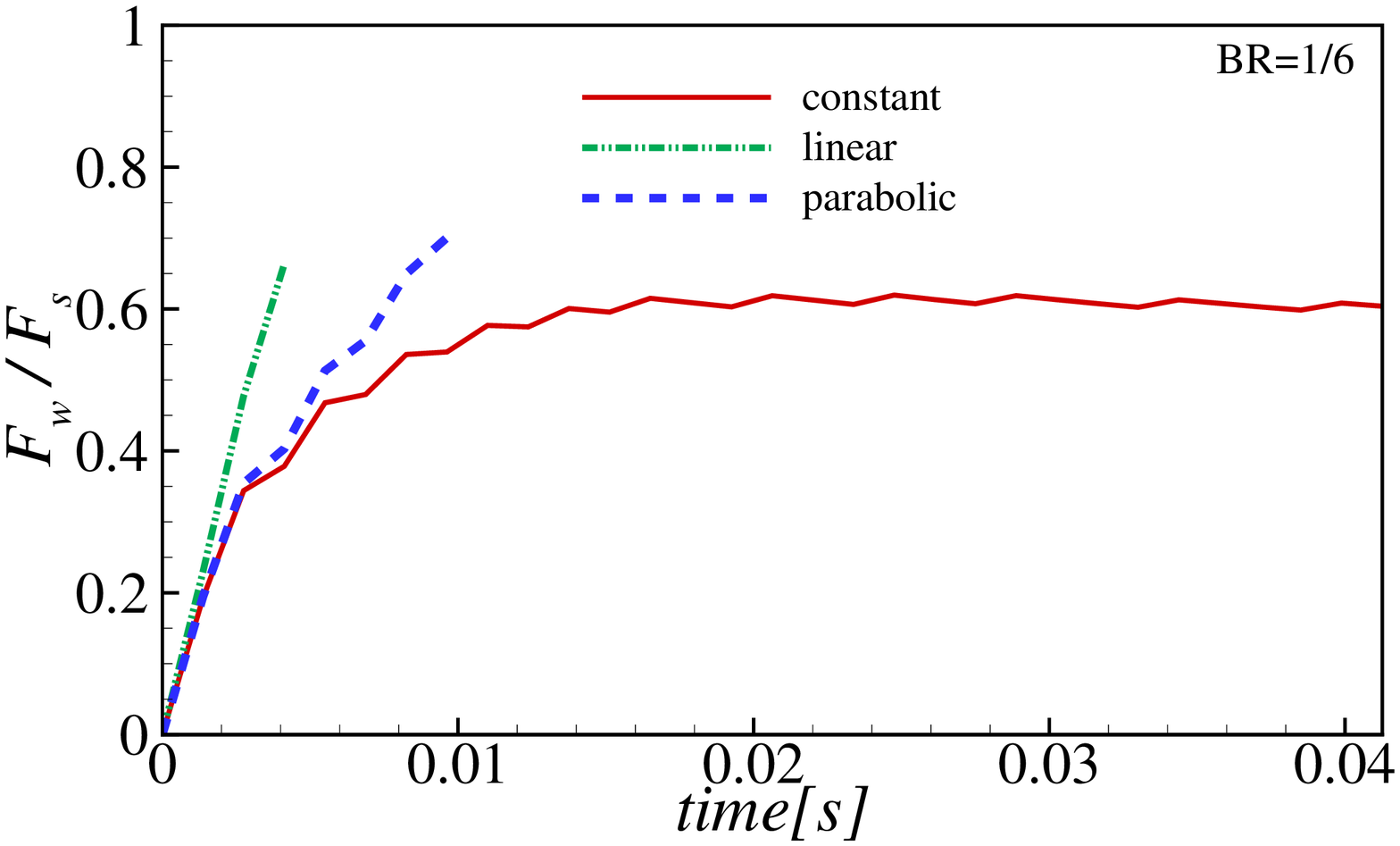}}

\subfloat[]{\label{fig: Forces_water_BR1.5_z0=0}\includegraphics[width=0.33\linewidth]{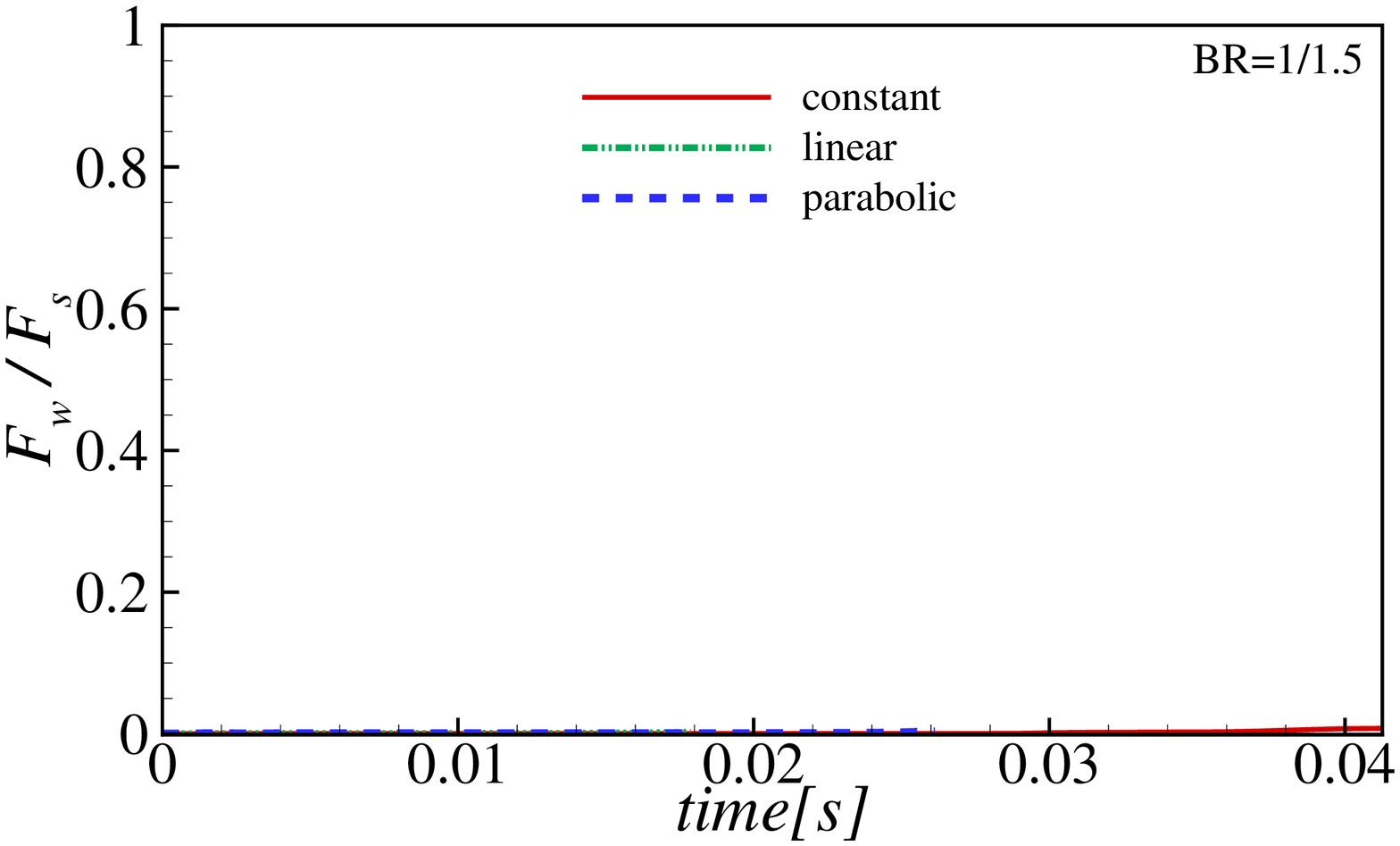}}
\subfloat[]{\label{fig: Forces_water_BR3_z0=0}\includegraphics[width=0.33\linewidth]{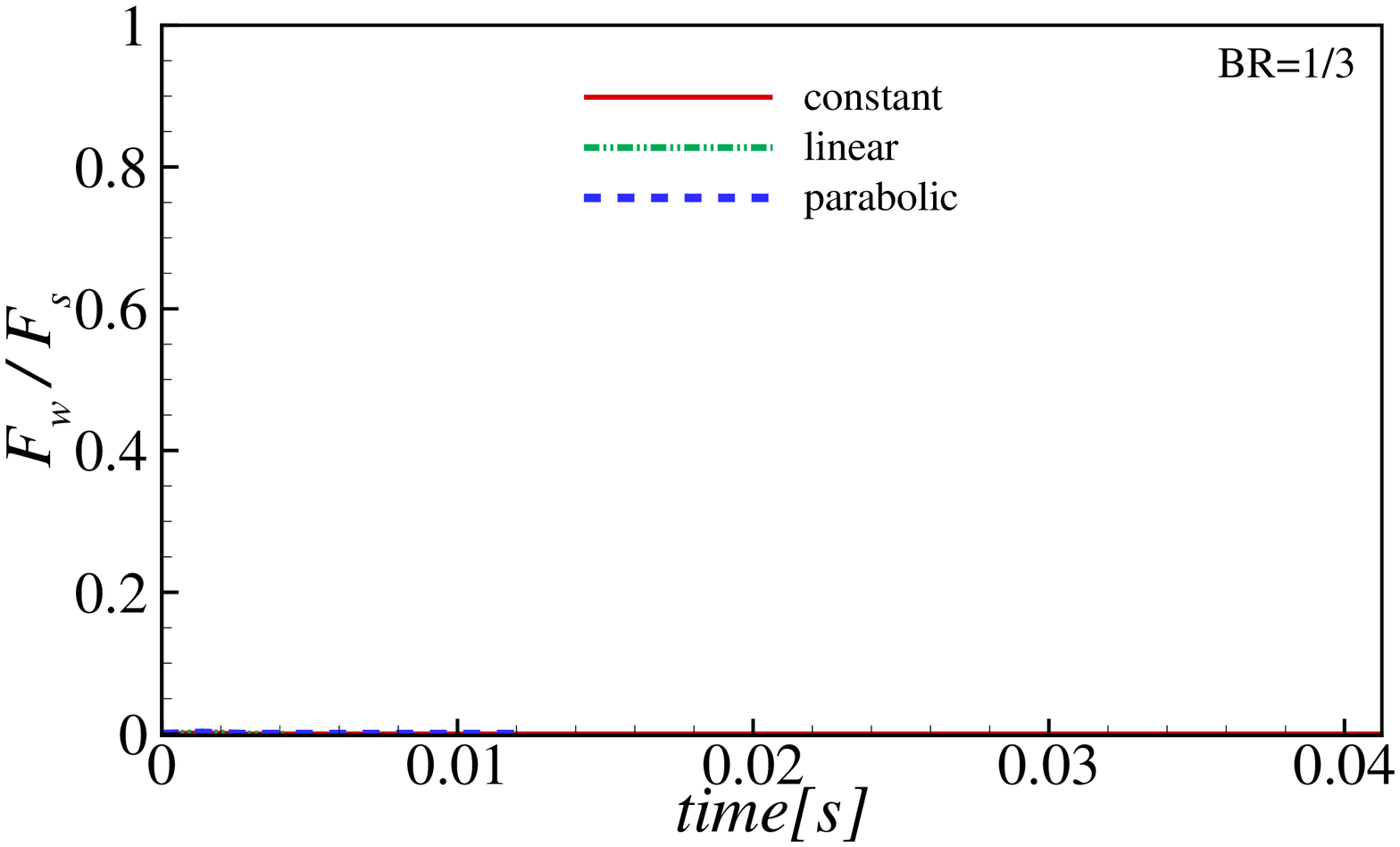}}
\subfloat[]{\label{fig: Forces_water_BR6_z0=0}\includegraphics[width=0.33\linewidth]{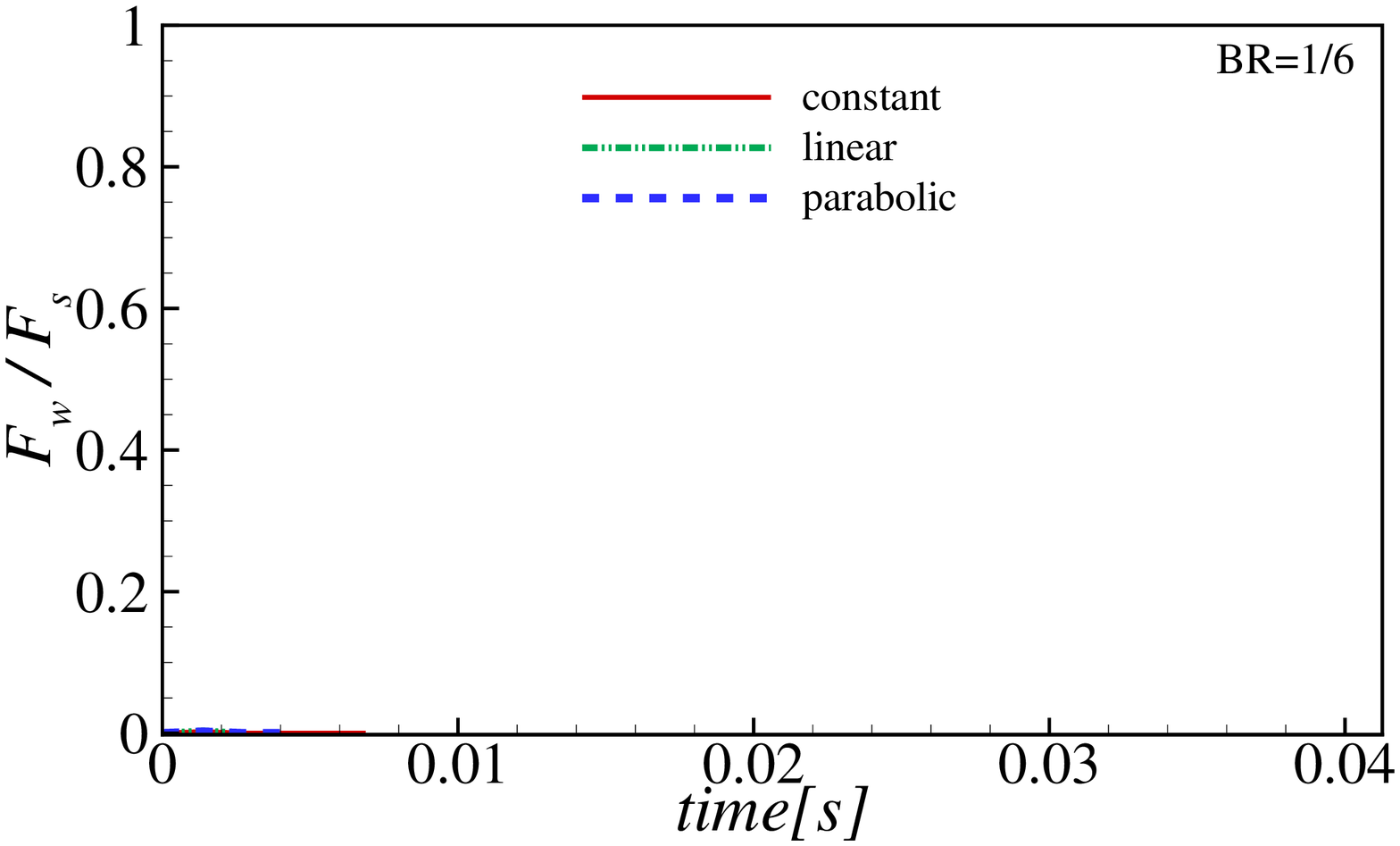}}

\caption{Contribution of the friction drag to the total force sense by the inner cylinder for the cases in which $z_0=0$, for different blockage ratios and different imposed velocity profiles to the sliding cylinder.: (a)-(c) silicon-1000, (d)-(f) silicon-1, and (g)-(i) water.}
\label{fig:3}       % Give a unique label
\end{figure}

\begin{figure}[h!]
\centering
\subfloat[Water, BR=1/1.5.]{\label{fig:alphaW-BR1.5}\includegraphics[width=0.9\linewidth]{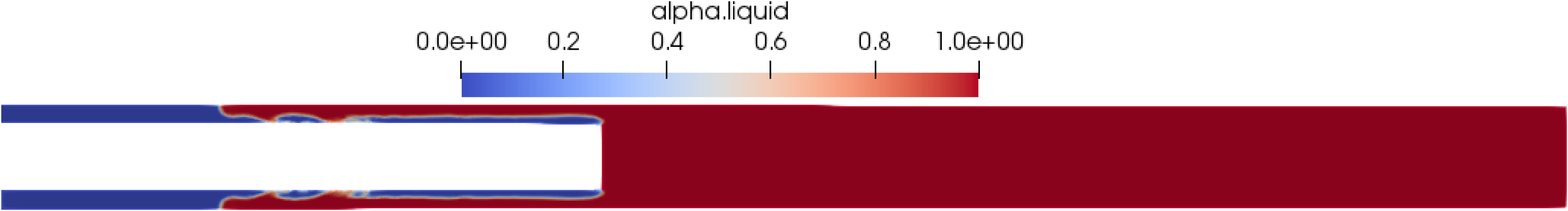}}\\
\subfloat[Water, BR=1/3.]{\label{fig:alphaW-BR3}\includegraphics[width=0.9\linewidth]{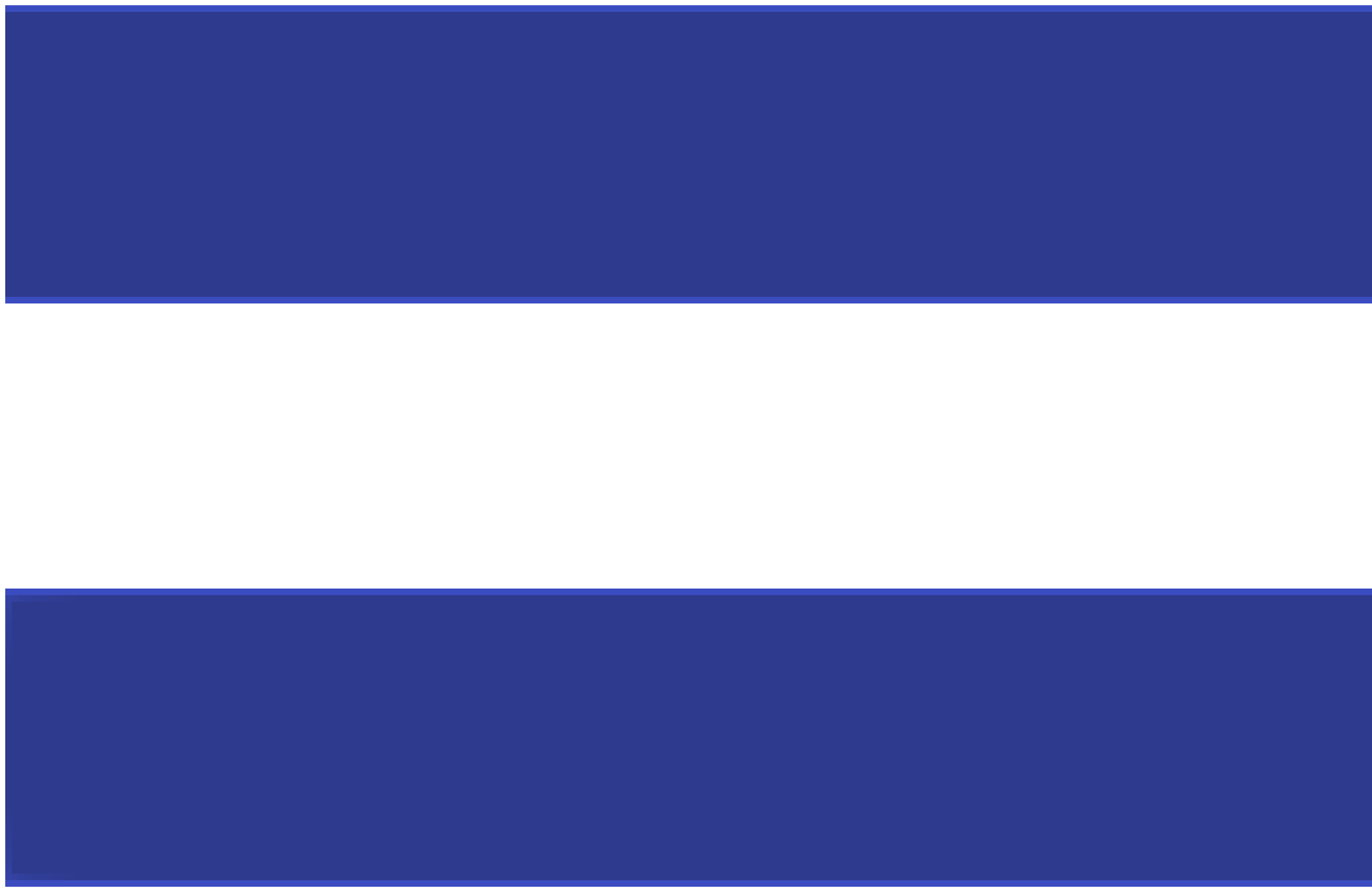}}\\
\subfloat[Water, BR=1/6.]{\label{fig:alphaW-BR6}\includegraphics[width=0.9\linewidth]{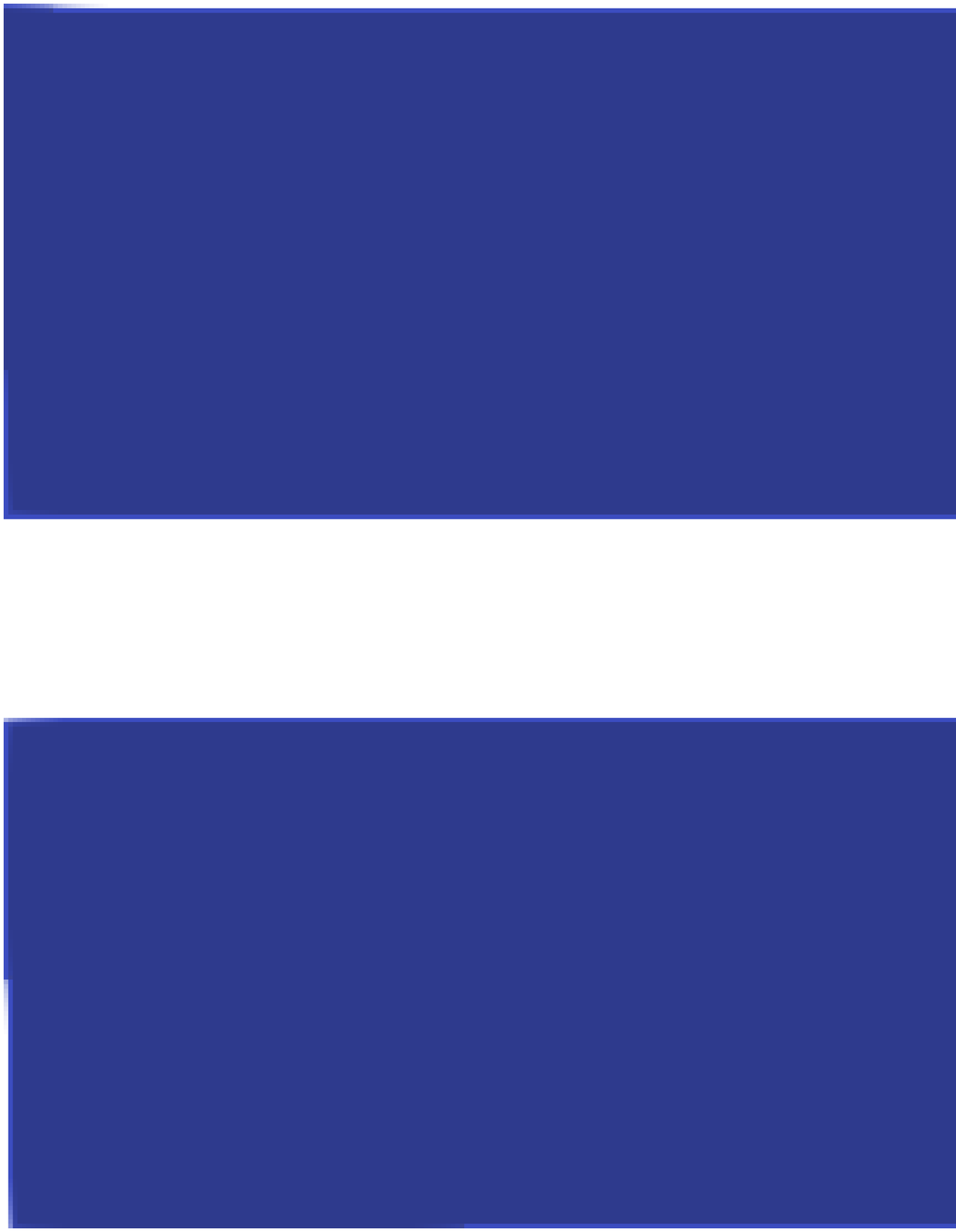}}\\
\subfloat[Silicon1000, BR=1/3.]{\label{fig:alphaWS1000}\includegraphics[width=0.9\linewidth]{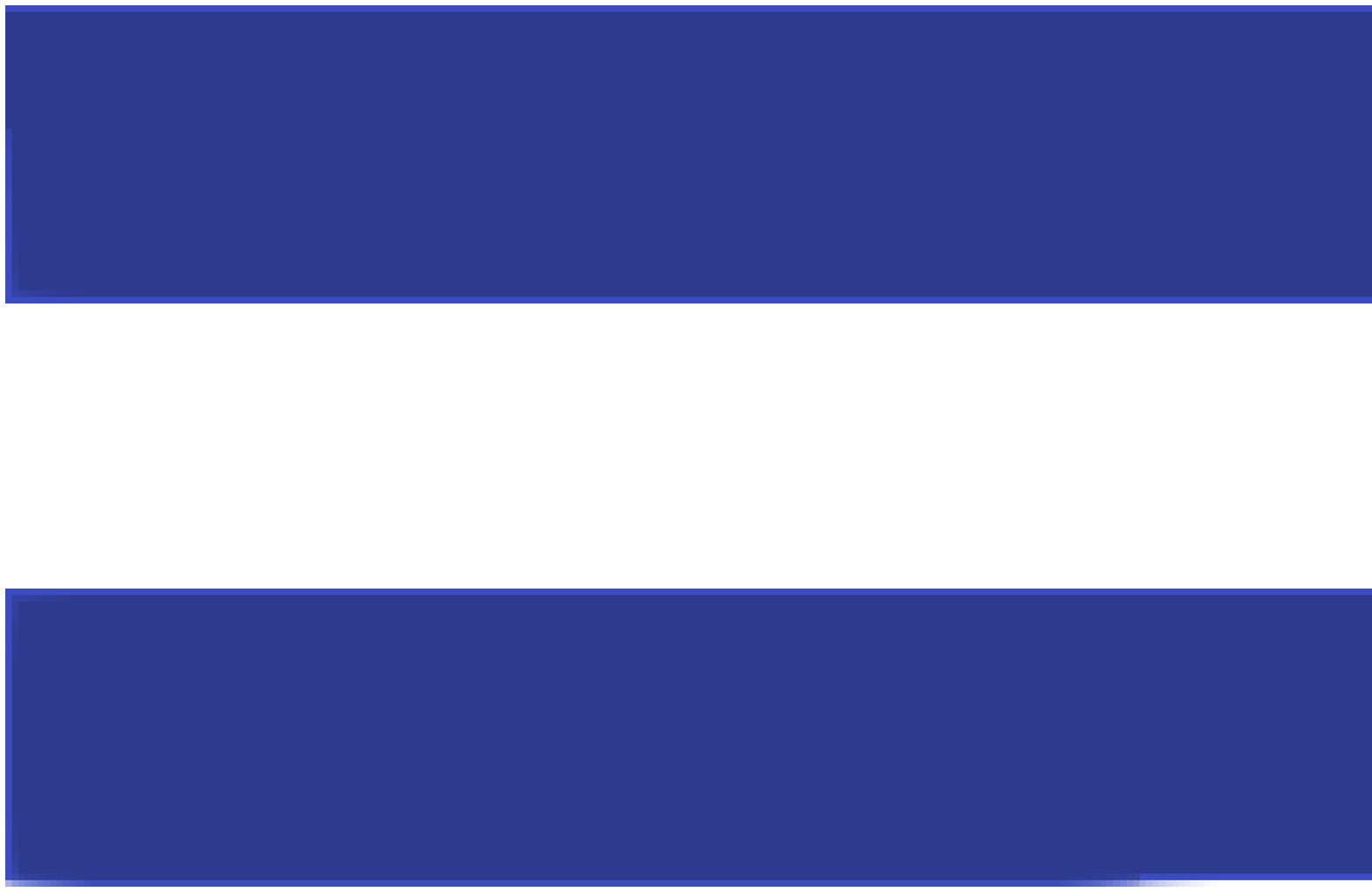}}
\caption{(a) to (c): Water splash when impactor moves in the case of a parabolic velocity  profile for the inner cylinder and for blockage ratios 1/1.5, 1/3 and 1/6 respectively. (d) Silicon-1000 ($z_0=0$) for the case of BR=1/3.}
 \label{fig:5}      % Give a unique label
\end{figure}

\begin{figure}[h!]
\centering

\subfloat[]{\label{fig: Forces_silicon_parabolicVelocity_BR1.5_z0=d}\includegraphics[width=0.5\linewidth]{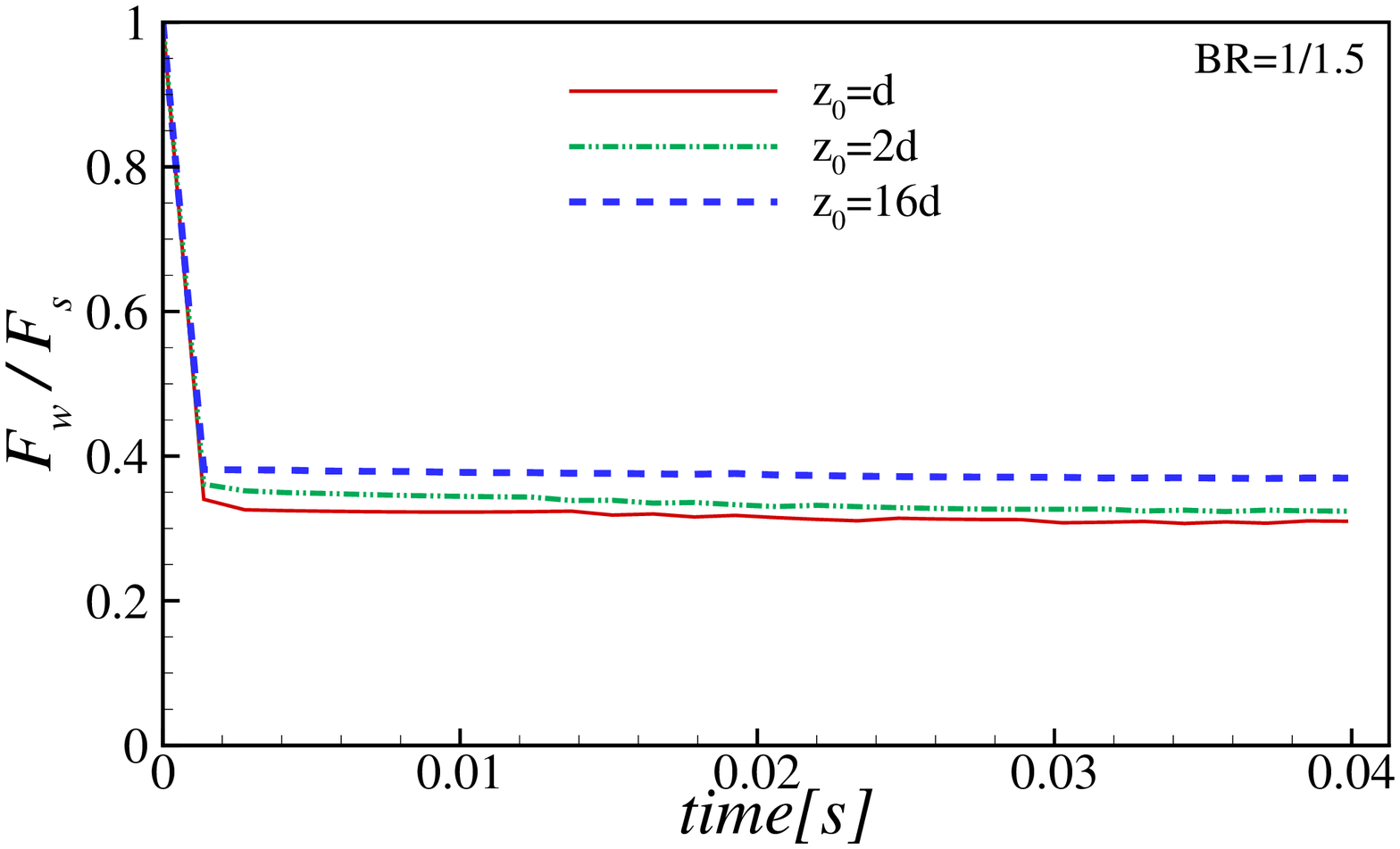}}
\subfloat[]{\label{fig: Forces_silicon_parabolicVelocity_BR1.5_z0=2d}\includegraphics[width=0.5\linewidth]{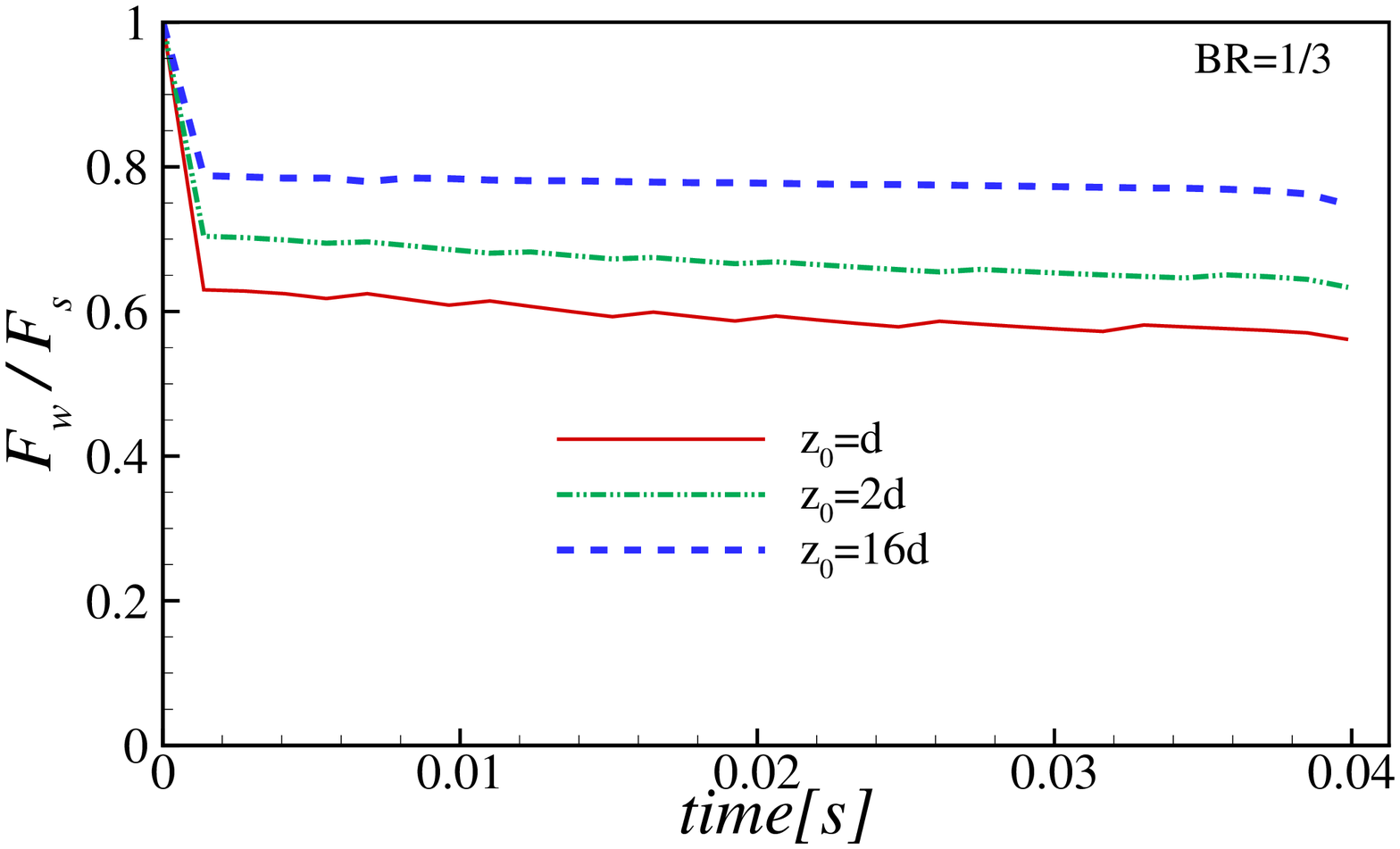}}

\subfloat[]{\label{fig: Forces_silicon_parabolicVelocity_BR1.5_z0=d2}\includegraphics[width=0.5\linewidth]{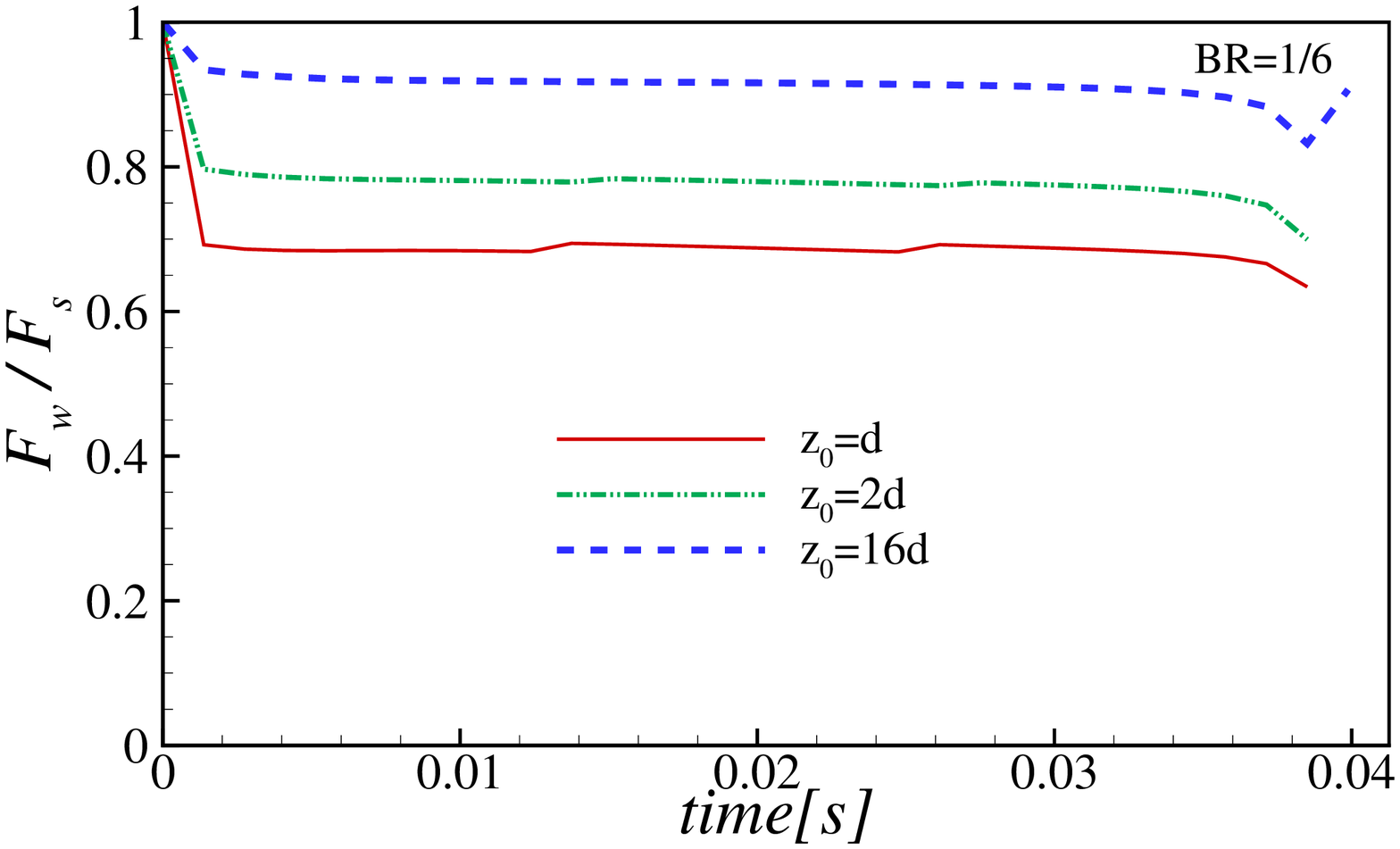}}

\caption{Normalized force for silicon-1000 and imposed parabolic velocity at different blockage ratio and different $z_0\neq0$.}
\label{fig:6}       % Give a unique label
\end{figure}

\begin{figure}[h!]
\centering
\subfloat[]{\label{fig:alphaS1000d2BR1.5}\includegraphics[width=\linewidth]{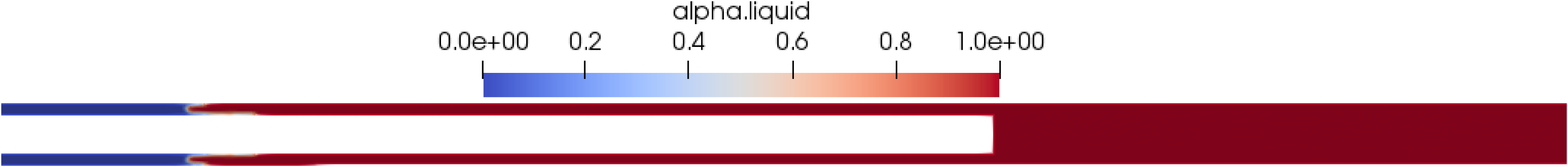}}

\subfloat[]{\label{fig:alphaS1000d2BR3}\includegraphics[width=\linewidth]{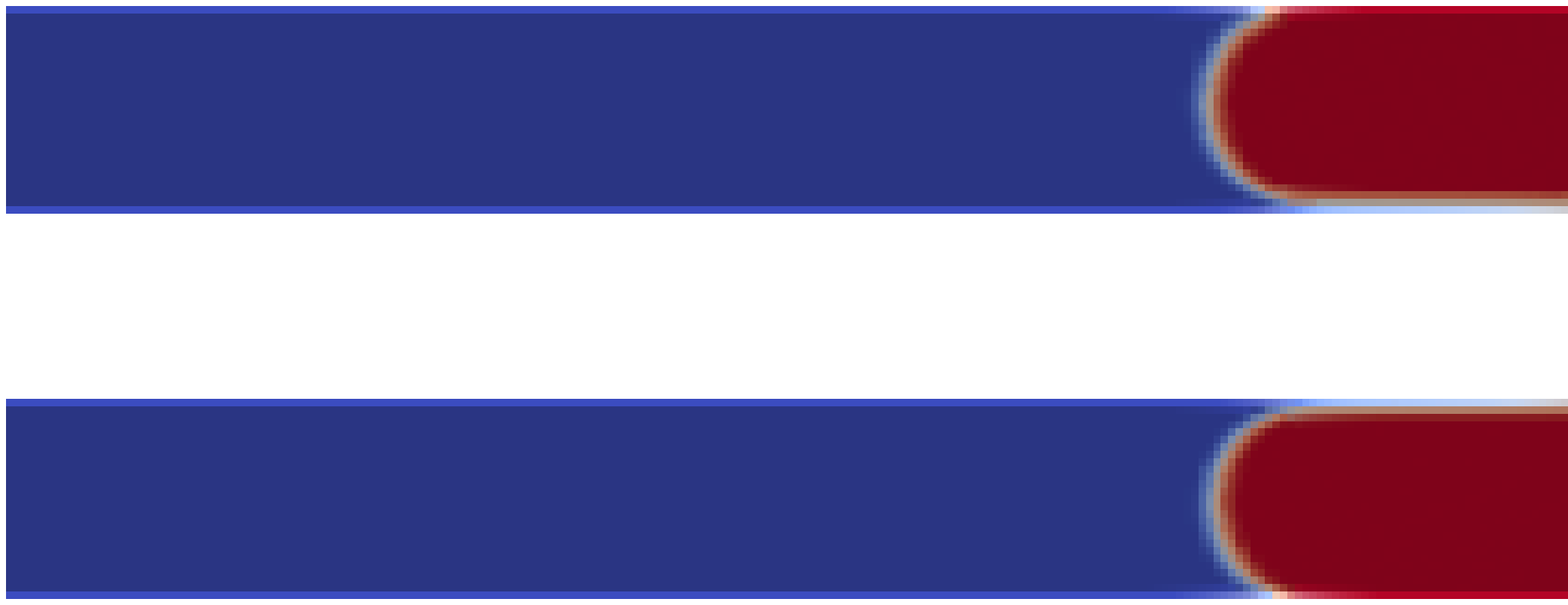}}

\subfloat[]{\label{fig:alphaS1000d2BR6}\includegraphics[width=\linewidth]{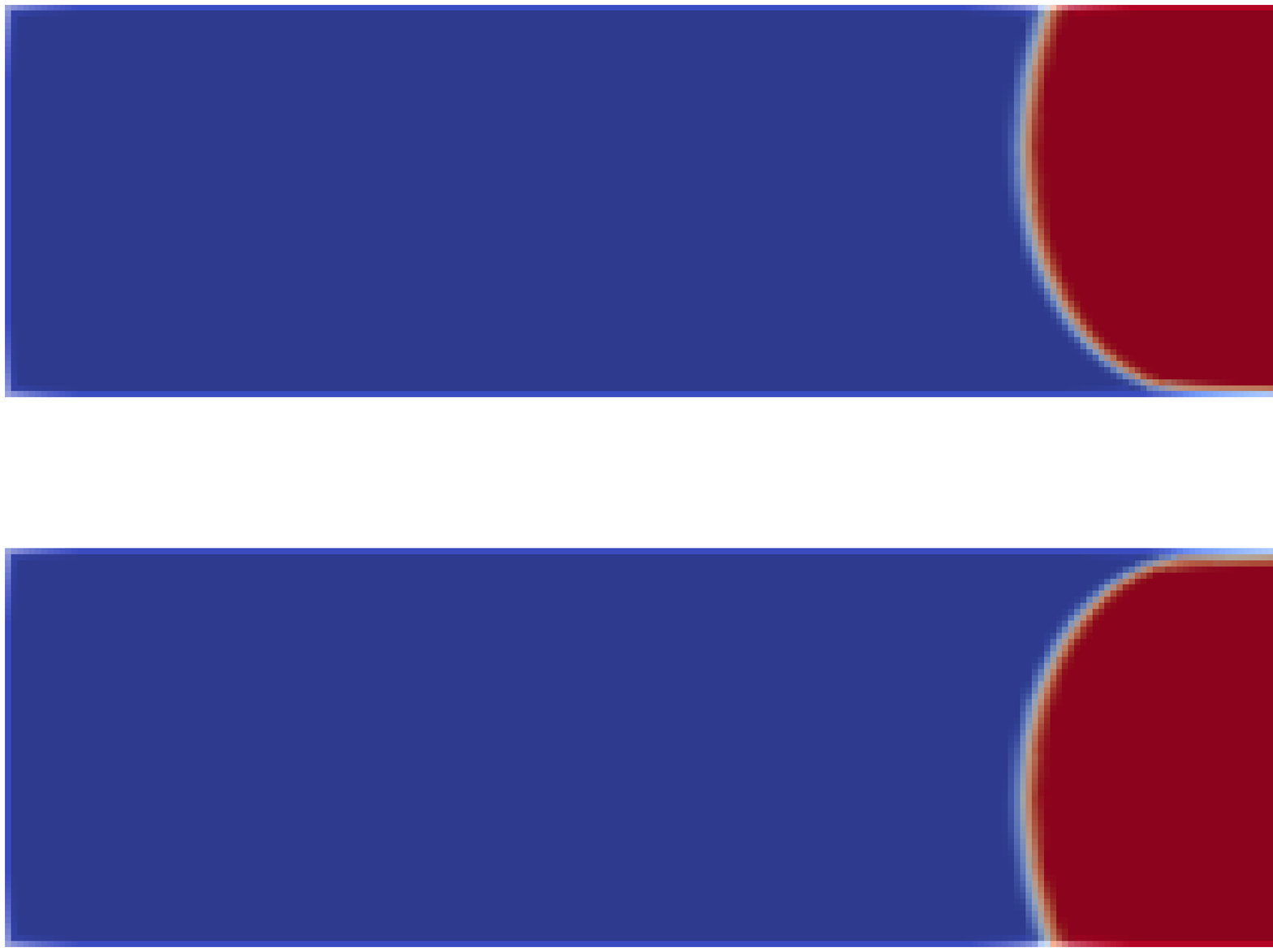}}

\caption{Influence of blockage ratio on the contact between the inner cylinder and the liquid for the case of using silicon-1000, a parabolic velocity profile and $z_{0}$=16d: (a) BR=1/1.5, (b) BR=1/3 and (c) BR=1/6.}
 \label{fig:7}      % Give a unique label
\end{figure}

\section{Results and discussion}
\subsection{Analysis of the forces}
Let's consider that a motor would be controlling the movement of the inner cylinder and the response of the fluid would be measured experimentally by means of a piezo-electric 1-component force sensor installed at the inner cylinder. That force measured by the sensor ($F_{s}$) would be the sum of different contributions:

 \begin{equation} \label{eq:FSI}
F_{s}= F_{p} + F_{w}+F_{b},
\end{equation}

\noindent where $F_{p}$ would be the pressure drag, $F_{w}$ the friction drag and $F_{b}$ the buoyancy, all of them corresponding, obviously, to the inner cylinder. As experimentally it will be impossible to decouple the contribution of these forces to the total force measured by the sensor, if it is intended the penetroviscometer to provide the instantaneous viscosity, then the design should result in friction drags dominating over the other components, so that $F_{s}\approx F_{w}$. 

Fig. \ref{fig:3} shows the importance of the friction drag ($F_{w}$) with regards to the total force sense by the inner cylinder ($F_{s}$; Eq. \ref{eq:FSI}) for the three different fluids, the three different blockage ratios and the three different imposed velocity profiles to the sliding cylinder. Despite it cannot be observed, due to the normalisation, that the values of the friction drag are also lower for the cases with lower viscosities, it is indeed observed that the friction drag is negligible for viscosity values of the order of 1 mPa$\cdot$s, regardless the blockage ratio and the velocity profile of the cylinder (Figs. \ref{fig: Forces_water_BR1.5_z0=0}-\ref{fig: Forces_water_BR6_z0=0}). This is supported by Figs. \ref{fig:alphaW-BR1.5}-\ref{fig:alphaW-BR6} where the splash produced in the liquid avoid any contact with the lateral surface of the inner cylinder. Fig. \ref{fig:alphaWS1000} makes evident that an increment in the viscosity of the liquid results in an increase of the contact area of the liquid with the inner cylinder and, therefore, $F_{w}$ becomes more relevant. Despite $F_{w}$ increases with the viscosity and with smaller blockage ratios, none of the cases shown in Fig. \ref{fig:3} can be considered as useful for calculating the instantaneous viscosity from the measurement of $F_{s}$, since $\frac{F_{w}}{F_{s}}\ll 1$ in all cases.

In order to increase the value of $F_{w}$, it would be required to start the experiment with the tip of the inner cylinder submerged a distance $z_{0}>0$. Fig. \ref{fig:6} compares the friction and pressure drags for different values of $z_{0}$ and for the different blockage ratios, but just for the case of imposing a parabolic profile to the sliding cylinder and for the silicon-1000. It can be observed that increasing $z_{0}$ up to a value of $16d$ results in a friction drag dominant over the pressure drag and buoyancy, for small blockage ratios (BR$\leq$1/3). Fig. \ref{fig:7} shows that it would be preferable to use a BR as small as possible, in order to to minimize the interaction with the outer wall and assume $F_s \approx F_w$. It can also be observed that, in these later cases, there is a mass added effect when the experiment is started, which results in an initial peak in the friction force that vanishes after $\sim1.3$ ms.

\subsection{Inertial artefacts} 

When measuring with the penetroviscometer, inertia can interfere in the measurement of $F_{w}$, depending on the viscosity of the fluid sample and the geometry (i.e. $d$, $D$, BR and $z_{0}$), which may potentially lead to artificial results for the instantaneous viscosity (Section \ref{Section:InstVisco}). Figures \ref{fig:alphaW-BR1.5}-\ref{fig:alphaW-BR6} show the problems of measuring the instantaneous viscosity with the penetroviscometer for water-like fluids when $z_{0}=0$ at any BR, due to the lack of lateral contact between the sample and the inner cylinder. Figure \ref{fig:6} showed that increasing the viscosity, the value of $z_{0}$ and BR will provide us $F_{s}\approx F_{w}$; however there is inertia problems start at $t=0$ s due to the added mass effect.  

An added mass force is created when the mass of fluid surrounding a body is suddenly accelerated or decelerated. Unavoidably, additional fluid forces will act on the surfaces in contact with the fluid and the measurement of $F_{s}$ will be affected by these forces. Therefore, it is required to defined when this inertial artifacts occur and define the range of reliability for the penetroviscometer to determine the instantaneous viscosity. Therefore, this added mass force will only appear at the beginning in those cases in which the sliding cylinder is already submerged in the  $z_{0}\neq 0$, and at the end of the experiment in those cases in which the velocity profile of the sliding cylinder is not constant. For the computation of this force, only the volume of the sliding cylinder submerged into the liquid will be considered, as it is around three order of magnitude denser than air:

\begin{equation}\label{eq:addeddrag}
F_{a}=m_{a} \frac{d}{dt}\left(v\left(t\right)-v_{l}\left(t\right)\right),
\end{equation} 
\noindent where $m_{a}=\rho_{l}\frac{\pi d^2}{4}z\left(t\right)$ is the added mass  and represents the equivalent added mass of the entire flow field about the accelerating/decelerating body, $v \left(t\right)$ is the velocity of the inner cylinder and $v_{l}\left(t\right)$ is the velocity of the liquid surrounding it. It is also important to remember that $z\left(t\right)=z_{0}+\int_{0}^{t}v\left(t\right)dt$, and therefore $z_{0}$ is an amplifier parameter for the added mass force. Looking at Eq.\ref{eq:addeddrag} it can be observed that both, at the very beginning, when $v_{l}\approx 0$ and $v(t)$ is at its maximum, and at the very end of the experiment, when $v_{l}$ is at its maximum and $v(t)\approx 0$, result in $|F_{a}|\gg F_{w}$.  
 
\subsection{Instantaneous viscosity}\label{Section:InstVisco}
%\textcolor{blue}{COMMENT: Ahmad, we need to explain how to calculate the instantaneous viscosity. Please read thoroughly this new part of the paper and let me know if it makes sense to you.}
In order to calculate the instantaneous viscosity $\eta\left(t\right)$, it is just required to compute the shear stress at the wall of the inner cylinder $\tau_{w}\left(t\right)$, the shear rate at the wall of the inner cylinder $\dot\gamma_{w}\left(t\right)$, and divide one by the other as in Eq. \ref{Eq:InstVisco}:
\begin{equation}
\eta\left(t\right)=\frac{\tau_{w}\left(t\right)}{\dot\gamma_{w}\left(t\right)}
\label{Eq:InstVisco}
\end{equation}

Similarly to the calculation made in a rotational rheometer, in the penetroviscometer $\tau_{w}\left(t\right)$ is proportional to the force measured by the sensor $F_{s}\left(t\right)$ (Eq. \ref{Eq:shearstress}):

\begin{equation}
\tau_{w}\left(t\right)=\frac{F_{s}\left(t\right)}{A\left(t\right)}=\frac{F_{s}\left(t\right)}{\left[z_{0}+z\left(t\right)\right]\pi d}=\frac{F_{s}\left(t\right)}{\left[z_{0}+\int^{t}_{0}v\left(t\right)dt\right]\pi d},
\label{Eq:shearstress}
\end{equation} 

\noindent where the constant of proportionality ($\frac{1}{\left[z_{0}+\int^{t}_{0}v\left(t\right)dt\right]\pi d}$) depends exclusively on the experimental parameters $d$, $v\left(t\right)$ and $z_{0}$. 

%\textcolor{blue}{COMMENT: Ahmad, we need a better quality image for Fig. \ref{fig:velocityProfilesLiquid}.}\\
\begin{figure}[t!]
\centering
%\subfloat[$t=0$ ms]{\label{fig:velocprofile_t=0}\includegraphics[width=0.15\linewidth]{velocprofile_t=0ms.png}}\hfill
\subfloat[$t=2.5$ ms]{\label{fig:velocprofile_t=0}\includegraphics[width=0.49\linewidth]{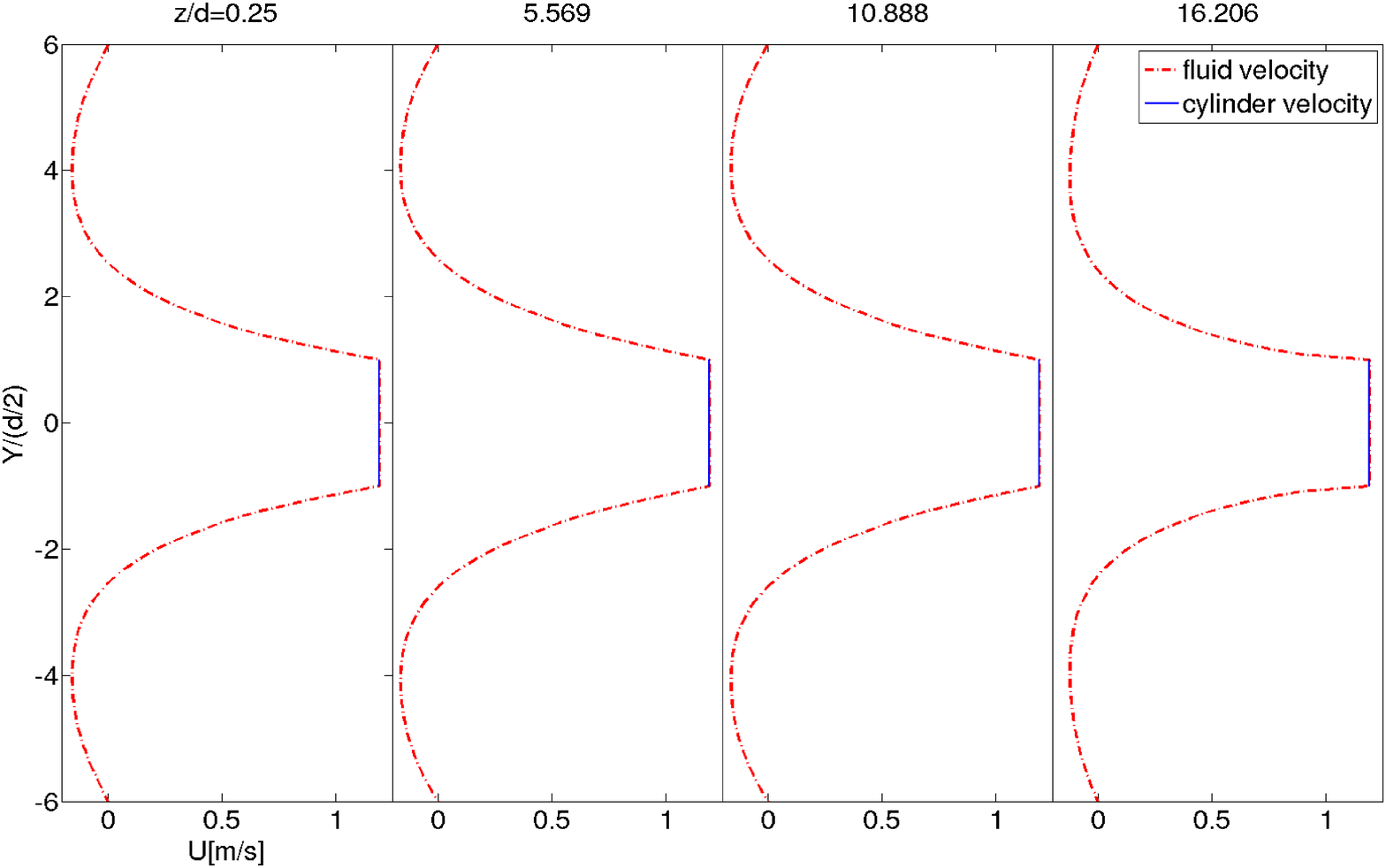}}\hfill
\subfloat[$t=5$ ms]{\label{fig:velocprofile_t=0}\includegraphics[width=0.49\linewidth]{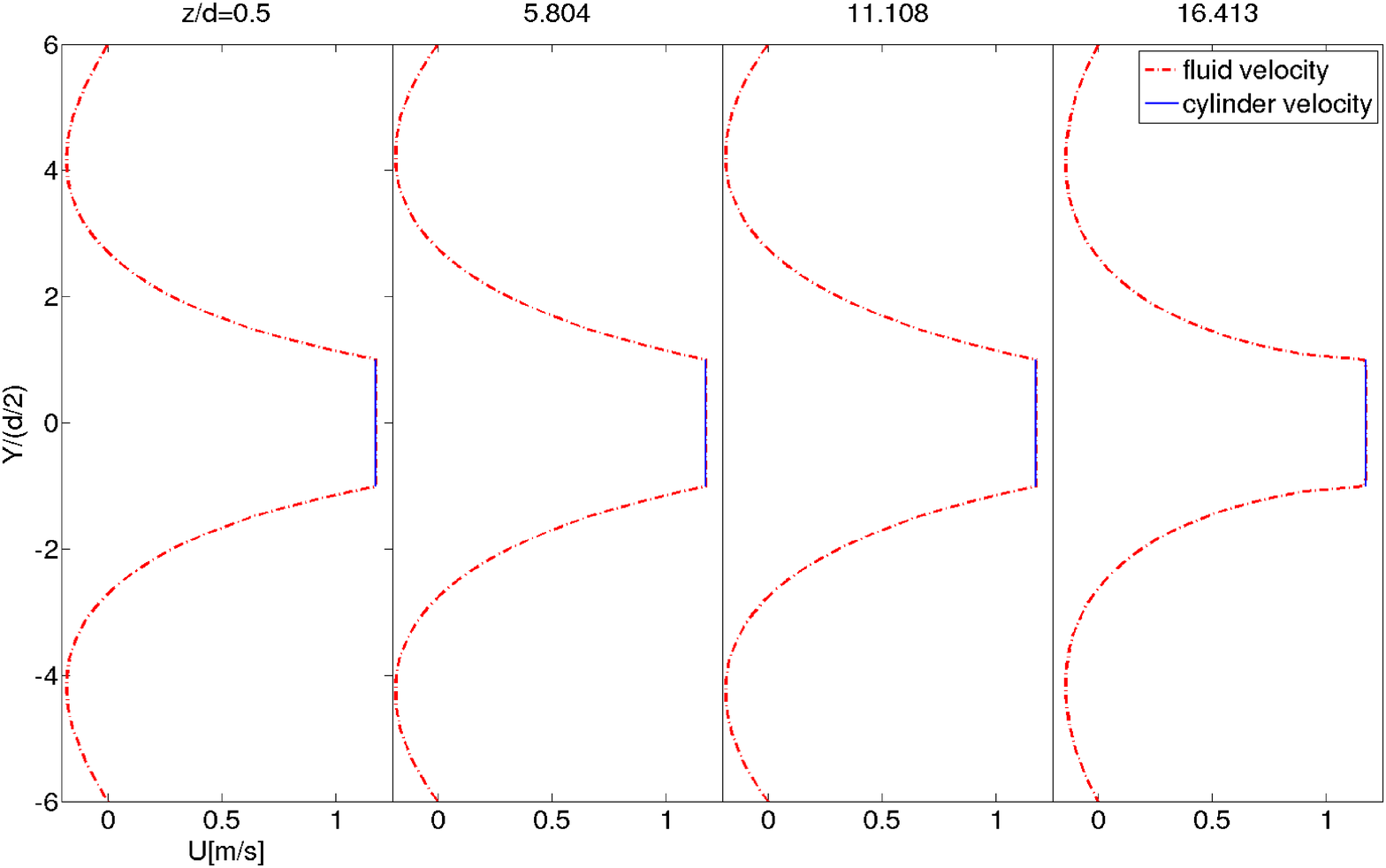}}\hfill

\subfloat[$t=10$ ms]{\label{fig:velocprofile_t=0}\includegraphics[width=0.49\linewidth]{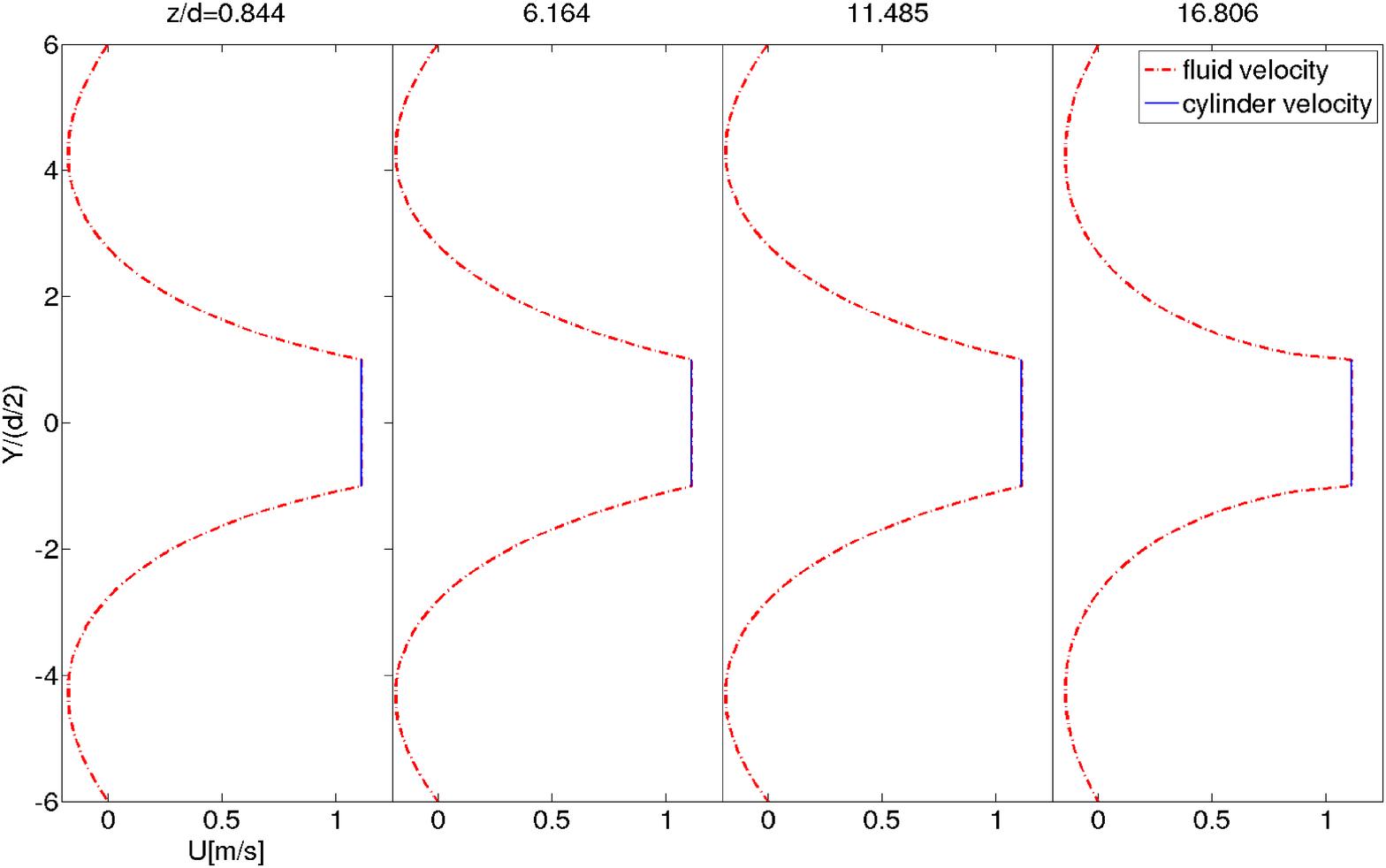}}\hfill
\subfloat[$t=20$ ms]{\label{fig:velocprofile_t=0}\includegraphics[width=0.49\linewidth]{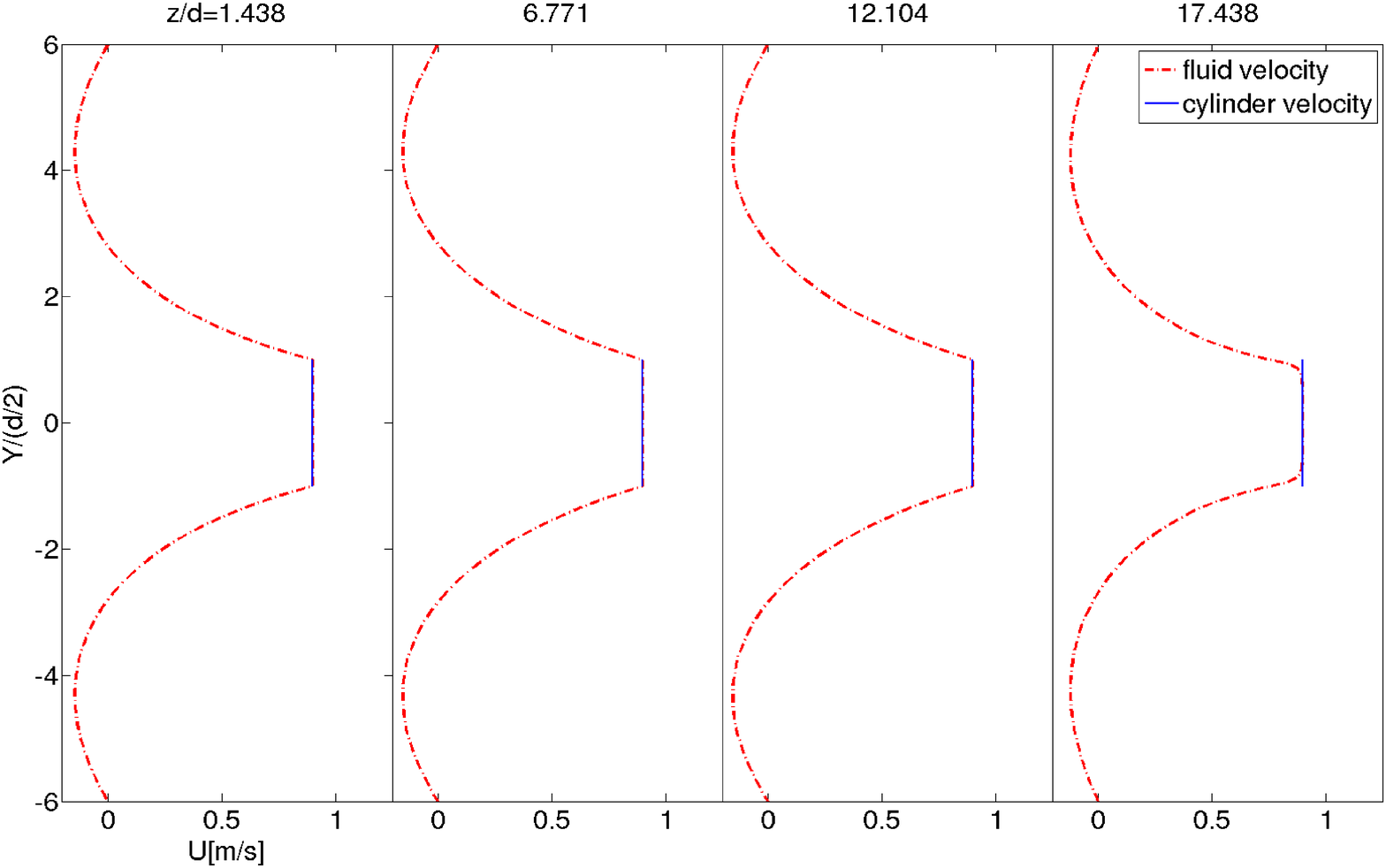}}\hfill

\subfloat[$t=40$ ms]{\label{fig:velocprofile_t=0}\includegraphics[width=0.49\linewidth]{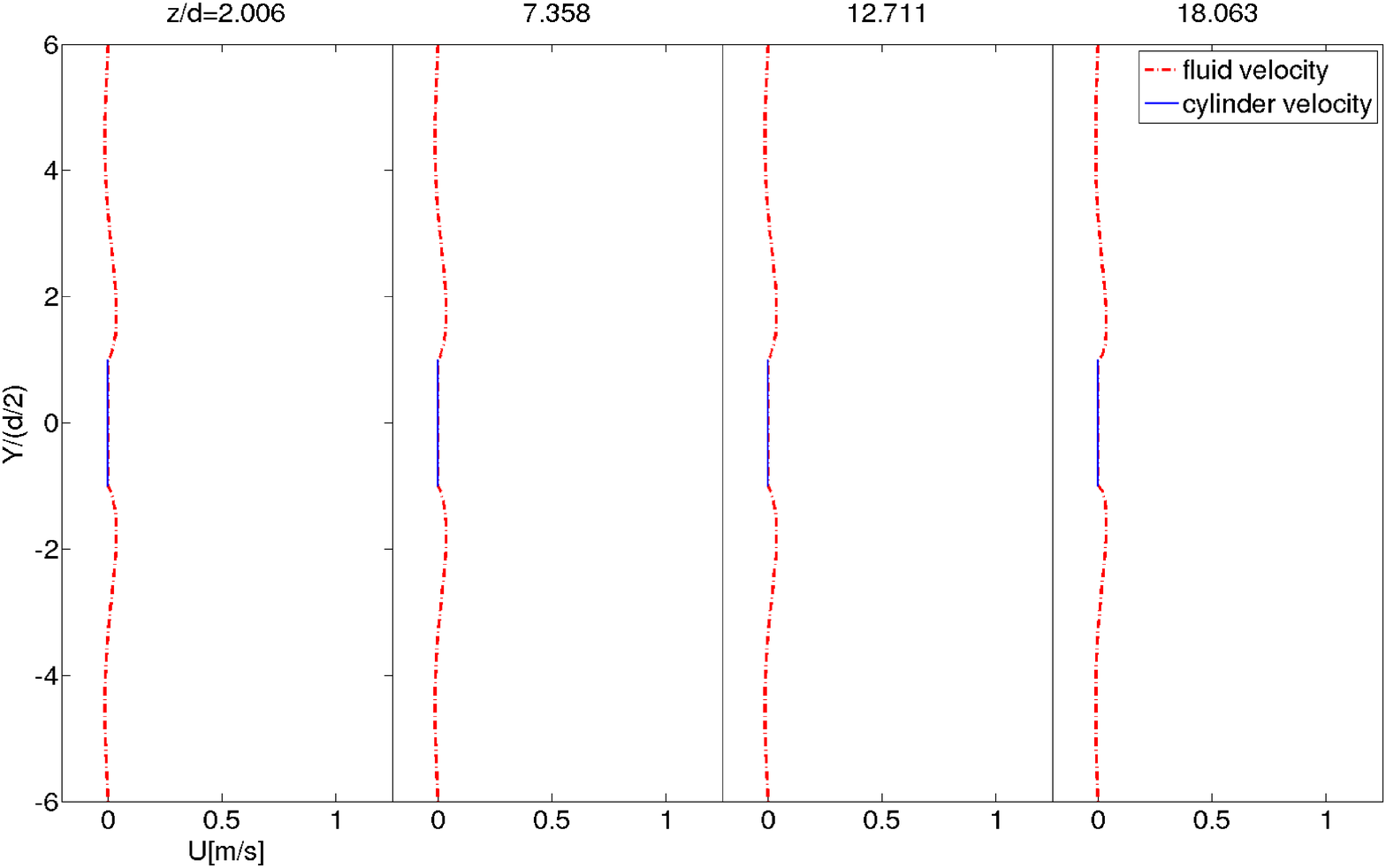}}\hfill
\caption{Fluid velocity profile at different z-positions and different instant of time for the case of using silicon-1000, a parabolic velocity profile, $z_{0}=16d$ and BR$=1/6$.}
 \label{fig:velocityProfilesLiquid}      % Give a unique label
\end{figure}

\indent The same philosophy is applied to the shear rate, which will also be a function of experimental parameters and the imposed velocity to the inner cylinder. Fig. \ref{fig:velocityProfilesLiquid} shows the velocity profile within the fluid between at different time steps and at different z-positions in contact with the inner cylinder. It can be observed that the velocity profile is not dependent of the z-position, but only depends on the radial position $r$ and it follows a quadratic expression ($v_{f}=a\left(r-\frac{d}{2}\right)^{2} + b\left(r-\frac{d}{2}\right) +c$). This velocity profile must satisfy the non-slip condition at both walls: $v_{f}\left(r=\frac{d}{2},t\right)=-v\left(t\right)$ and $v_{f}\left(r=\frac{D}{2},t\right)=0$; moreover, as the volume ($Q$) of fluid moved by the tip of the inner cylinder must be preserved, the following conservative condition must be also satisfied:

\begin{equation}
Q=v\left(t\right)\pi d^{2}/4=\int^{D/2}_{d/2} v_{f}\left(r\right)2\pi r dr. 
\label{Eq:conservQ}
\end{equation} 

\noindent In this way the following linear system of equations must be solved out in order to have the coefficients $a$, $b$ and $c$ of the velocity profile of the liquid:

\begin{equation} \label{eq:system}
\begin{bmatrix} 
\left(\frac{d}{2}\right)^{2} & \frac{d}{2} & 1 \\ 
\left(\frac{D}{2}\right)^{2} & \frac{D}{2} & 1 \\ 
\frac{1}{4}\left[\left(\frac{D}{2}\right)^{4}-\left(\frac{d}{2}\right)^{4}\right] & \frac{1}{3}\left[\left(\frac{D}{2}\right)^{3}-\left(\frac{d}{2}\right)^{3}\right]  & \frac{1}{2}\left[\left(\frac{D}{2}\right)^{2}-\left(\frac{d}{2}\right)^{2}\right]  \\ 
\end{bmatrix} 
\cdot
%\left[ 
\begin{bmatrix}
a \\
b \\
c
\end{bmatrix} 
%\right] 
=
%\left[
 \begin{bmatrix}
  -v\left(t\right) \\ 
  0 \\
  \frac{v\left(t\right)d^2}{8} 
  \end{bmatrix} ,
%  \right]
\end{equation}

\noindent which has the following solution:
\begin{equation}\label{eq:a}
a\left(t\right)=8v\left(t\right)\frac{D^2+Dd-5d^2}{\left(D^4-2D^3d+2Dd^3-d^4\right)},
\end{equation}
\begin{equation}\label{eq:b}
b\left(t\right)= -2v(t) \frac{\left( 3D^2+2Dd-11d^2 \right) }{\left( D-d \right) \left( D^2-d^2 \right)},\\
\end{equation}
\begin{equation}\label{eq:c}
c\left(t\right)=v(t).
\end{equation}

%\textcolor{blue}{COMMENT: Ahmad, we need to add a new figure (Fig. \ref{fig:coefficients}), which must show the numerical velocity profiles at different z-positions and at 5ms, for instance (solid line in red color). We should compare this numerical result with the analytical solution (dashed line in blue color), given by $v_{f}=a(d/2)^2 + b(d/2)+c$, where the coefficients $a$, $b$ and $c$ are given by Eqs. \ref{eq:a}-\ref{eq:c}. Please, have a look at the sketch that I depicted and let me know if you need me to explain...it may not be clear enough.}\\

\begin{figure}[t!]
\centering
\subfloat[$t=2.5$ ms]{\label{fig:velocprofileAnalyticalNumerical_t=1}\includegraphics[width=0.49\linewidth]{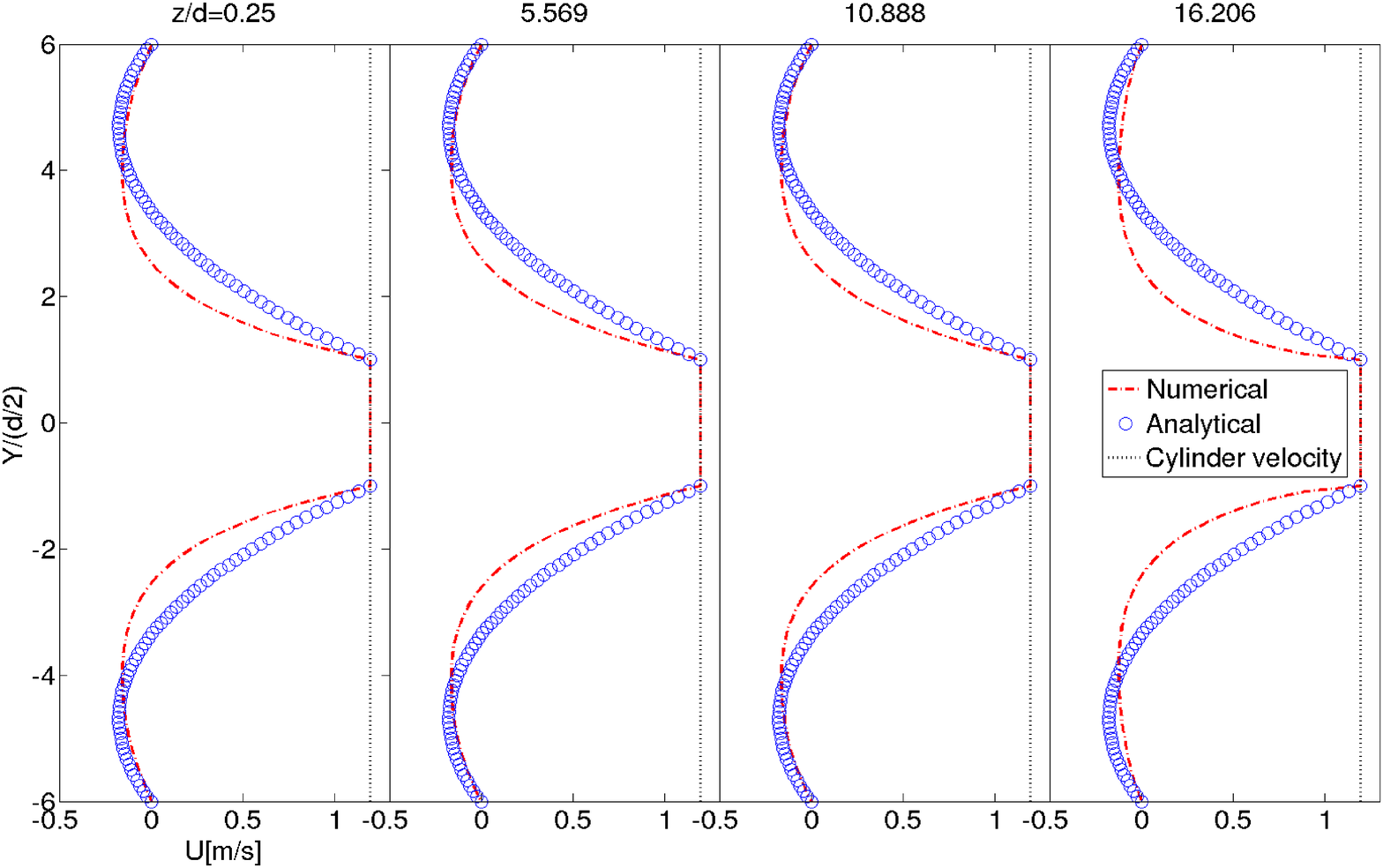}}\hfill
\subfloat[$t=5$ ms]{\label{fig:velocprofile_t=2}\includegraphics[width=0.49\linewidth]{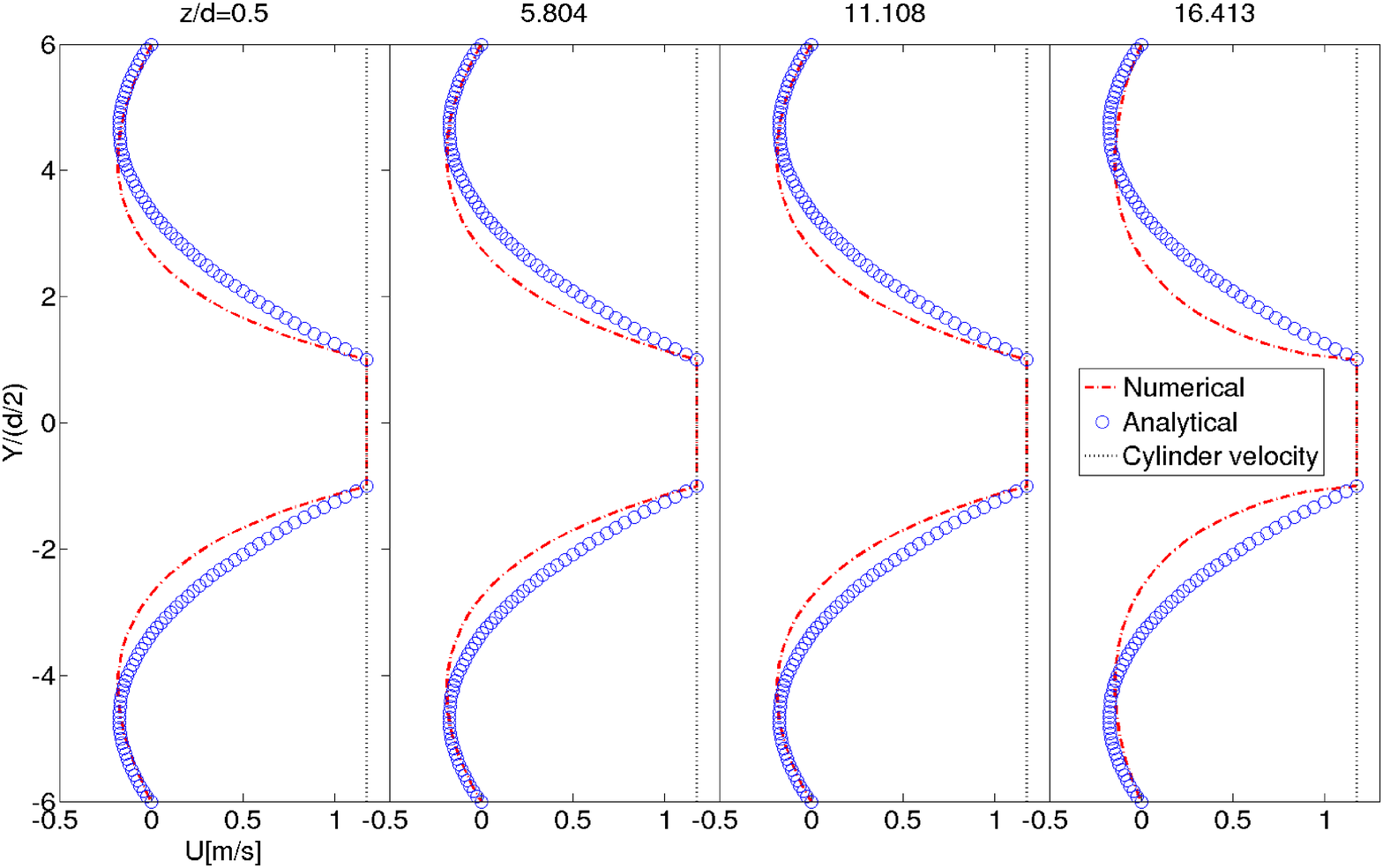}}\hfill

\subfloat[$t=10$ ms]{\label{fig:velocprofile_t=3}\includegraphics[width=0.49\linewidth]{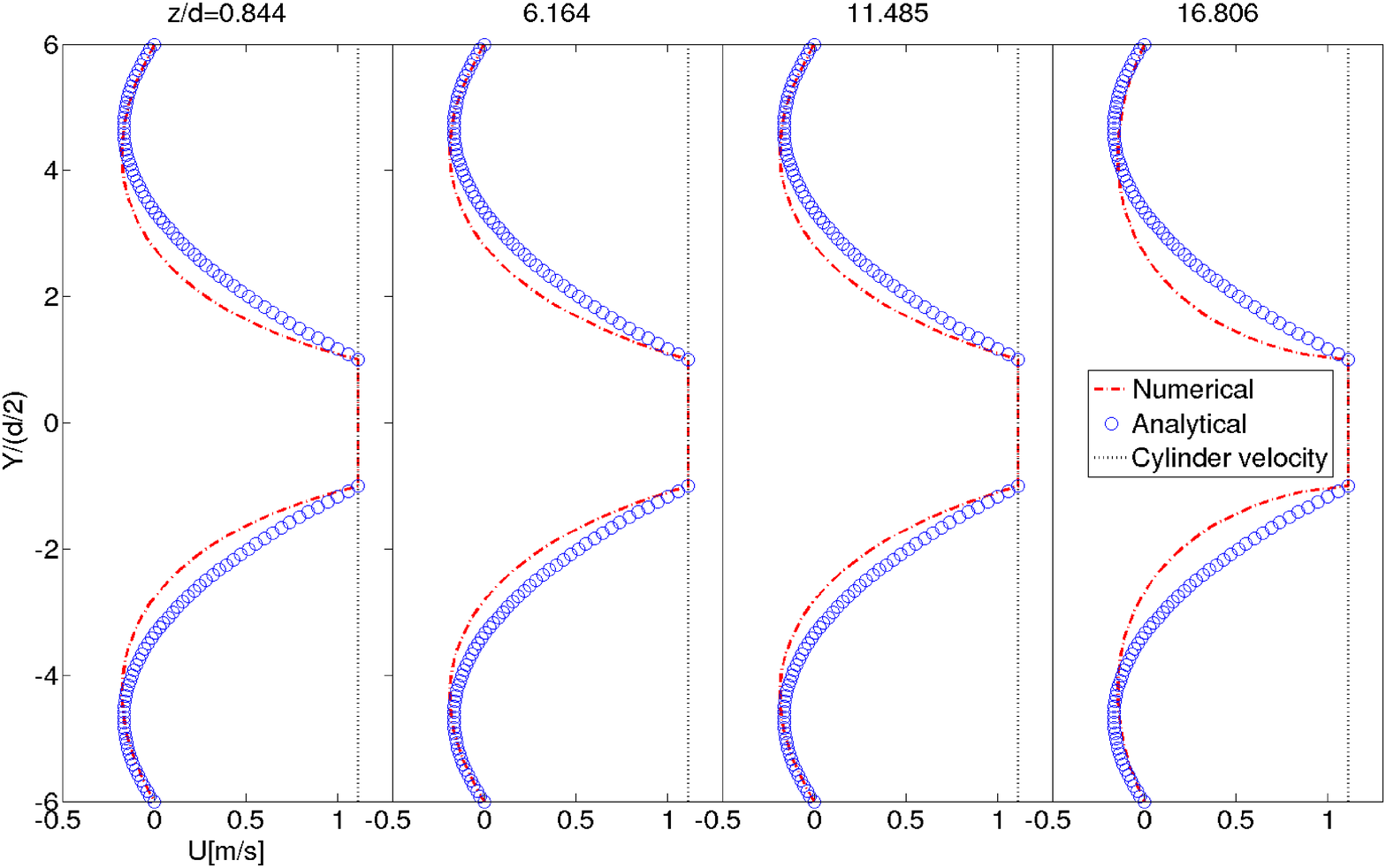}}\hfill
\subfloat[$t=20$ ms]{\label{fig:velocprofile_t=4}\includegraphics[width=0.49\linewidth]{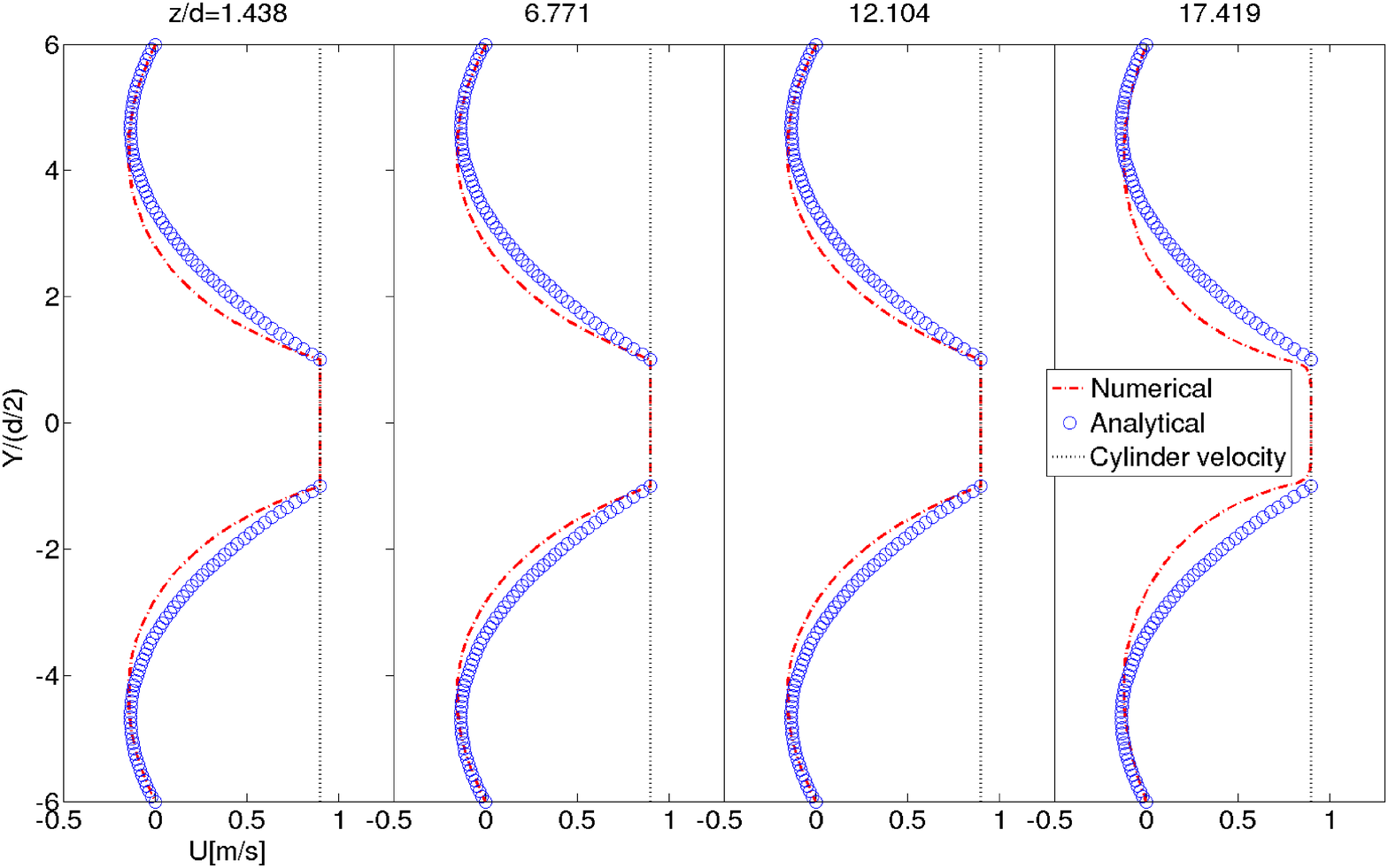}}\hfill

\subfloat[$t=40$ ms]{\label{fig:velocprofile_t=5}\includegraphics[width=0.49\linewidth]{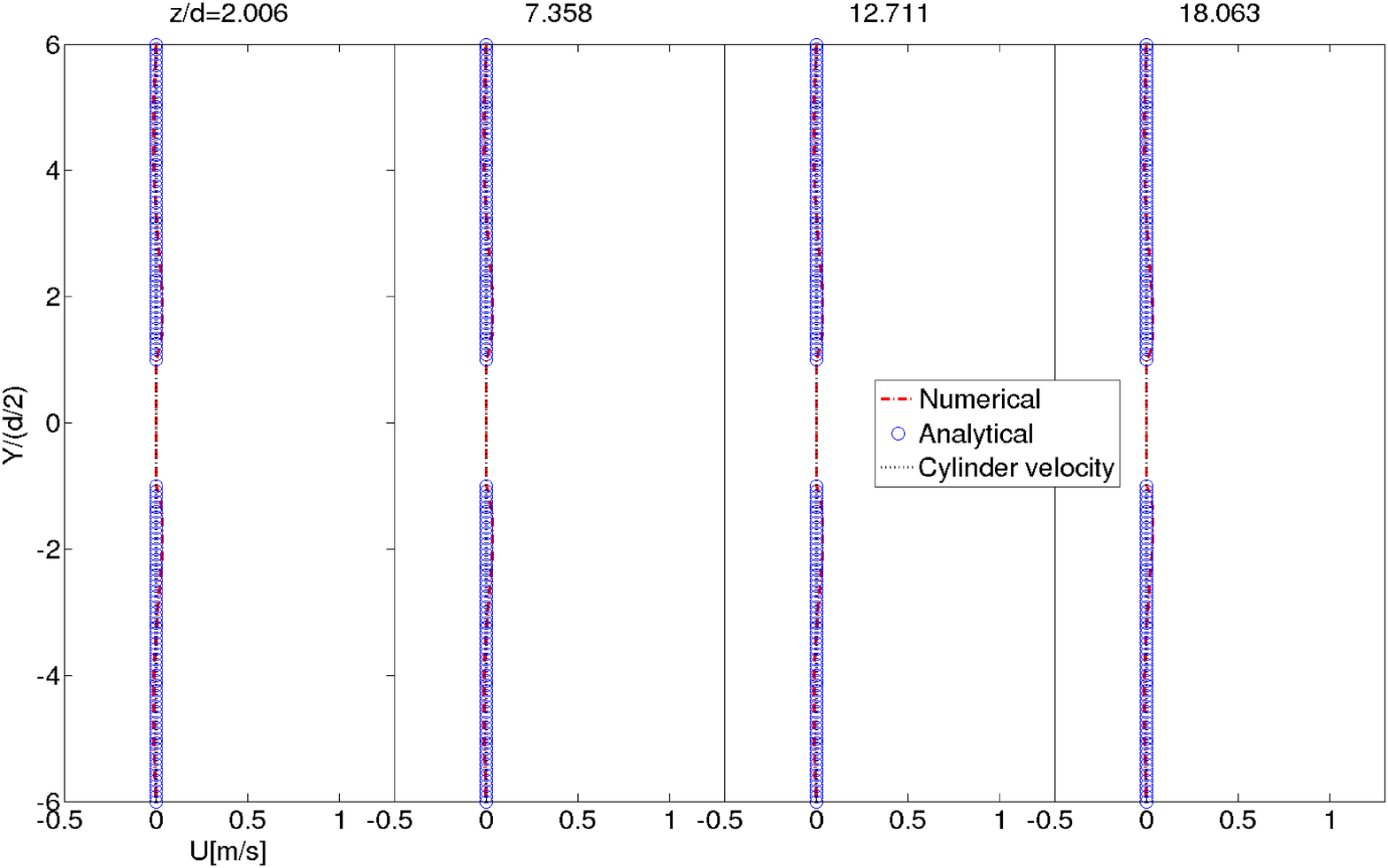}}\hfill

\caption{Comparison between the analytical and the numerical velocity profiles in the liquid at different z-positions and for the case of using silicon-1000, a parabolic velocity profile, $z_{0}=16d$ and BR$=1/6$. Instant of time $t=5$ ms.}
 \label{fig:velocityProfilesAnVsNum}      % Give a unique label
\end{figure}

\noindent These coefficients depend on time ($t$), but not on the z-position. Fig. \ref{fig:velocityProfilesAnVsNum} compares the numerical and the analytical solution for the velocity profile in the liquid contained between the two cylinders.\\ 

Once the velocity field in the fluid is known ($\vec{v_{f}}=v_{f}\left(t\right)\vec{e}_{z}$), the shear rate tensor can be calculated (Eq. \ref{eq:shearratetensor}).

\begin{equation} 
\bar{\bar{\dot\gamma}}=\frac{1}{2}\left(\nabla\vec{v_{f}}+\nabla\vec{v_{f}}^{T}\right)=
\left(\begin{matrix} 
0 & 0 & \frac{dv_{f}}{dr} \\ 
0&0& 0 \\ 
\frac{dv_{f}}{dr}  & 0 & 0\\ 
\end{matrix} \right)
\label{eq:shearratetensor}
\end{equation}

 Consequently, the shear rate at the wall of the inner cylinder is given by Eq. \ref{eq:shearratescalar}, which is only a function of the velocity of the inner cylinder and the dimensions of the concentric cylinders:

\begin{equation} \label{eq:shearratescalar}
\dot\gamma_{w}\left(t\right)=\left. \frac{dv_{f}\left(t\right)}{dr} \right|_{r=\frac{d}{2}} =b\left(t\right)=-\frac{2v(t) \left( 3D^2+2Dd-11d^2 \right) }{\left( D-d \right) \left( D^2-d^2 \right)},\\
\end{equation}

Thus, the instantaneous viscosity is therefore given by Eq. \ref{Eq:InstVisco}, which can be rewritten as follows:
\begin{equation}
\eta\left(t\right)=-\frac{\tau_{w}\left(t\right)}{\dot\gamma_{w}\left(t\right)}=\frac{\left( D-d \right) \left( D^2-d^2 \right) F_s(t)}{2v(t) \left( 3D^2+2Dd-11d^2 \right) \left[z_{0}+\int^{t}_{0}v\left(t\right)dt\right]\pi d}
\label{Eq:InstVisco2}
\end{equation}

Eq. \ref{Eq:InstVisco2} is also useful for defining the experimental setup for the penetroviscometer, as it contains all the involved experimental variables: $d$, $D$, $z_{0}$, $v$ and $F_{s}$. In this way, for example, one could get an estimation of the order of magnitude of the viscosity; then considering BR$\sim 1/6$, and defining the velocity of the inner cylinder, it is possible to determine the range of the force transducer.

\begin{figure}[h!]
\centering

%\subfloat[Numerical shear rate versus the analytical value.]{\label{fig: gamaDot_versus_b}\includegraphics[width=0.5\linewidth]{gamaDot_b.eps}}
%\subfloat[Viscosity versus analytical shear rate.]{\label{fig: viscosity_versus_b}\includegraphics[width=0.5\linewidth]{viscosity_b.eps}}

%\subfloat[$\left( \eta_{computed}-\eta_{silison-1000} \right)/ \eta_{silison-1000} \times 100$.]{\label{fig: viscosity_error}

\includegraphics[width=\linewidth]{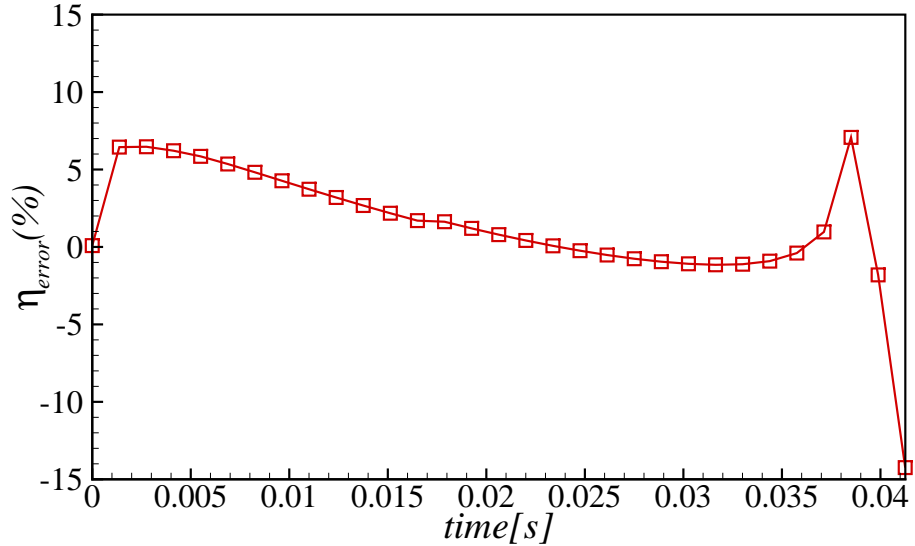}

\caption{Relative error in the calculation of the viscosity by means of the Eq. \ref{Eq:InstVisco2}.}
\label{fig:12}       % Give a unique label
\end{figure}

\section{Conclusion and future works}
Inspired by the penetroviscometer proposed by Bikerman\cite{BIKERMAN194875} more than 70 years ago, we have performed a numerical and analytical study to assess the usefulness of this kind of devices for measuring the instantaneous viscosity curve of shear thickening fluids under impact conditions. To do so, we have considered Newtonian fluids ranging from $10^{-3}$ to $10^{3}$ Pa$\cdot$s, three different blockage ratios (BR=1/1.5, BR=1/3 and BR=1/6), three different impact velocity profiles (constant, linear and parabolic) and three different initial positions for the inner cylinder (at the interface air-liquid $z_{0}=0$, and submerged into the fluid sample at $z_{0}=d$, $2d$ and $z_{0}=16d$). From the experimental point of view, the fluid sample must be in contact with the lateral area of the inner cylinder in order to compute the shear stress at the wall, and this is only accomplished for high viscous fluids ($\mu\gg1$ Pa$\cdot$s). Thus, it is expected to be useful for shear thickening fluids. Additionally, in order to calculate the instantaneous viscosity, the shear stress over the lateral area of the inner cylinder must dominate over the pressure at the tip; then, the results reported in this work recommended to start the experiments with the inner cylinder submerged into the fluid sample as much as possible and the gap between the two cylinder should be as large as possible. Moreover, inertial artefact may be an issue, particularly at the beginning and at the end of the experiment due to added mass effect; therefore, these experimental data should be discarded. We ended up with an analytical expression (Eq. \ref{Eq:InstVisco2}) that is able to provide the instantaneous viscosity based on geometric parameters ($z_{0}$, $d$ and $D$), the velocity of the inner cylinder ($v\left(t\right)$) and the measured force ($F_{s}$). For the case of silicon-1000, a parabolic profile for the inner cylinder,  $z_{0}=16d$, and BR=1/6, this device using Eq. \ref{Eq:InstVisco2} would be able to provide the instantaneous viscosity with an accuracy of $\approx 93\%$. In this sense, we can consider that these device may provide useful experimental data to help in the development of impact protective devices and new constitutive models accounting for the transient behavior of shear thickening fluids. However, in order to provide reliable data, the non-slip condition must be accomplished.

\bibliography{currobiblio}

\begin{thebibliography}{10}
\expandafter\ifx\csname url\endcsname\relax
  \def\url#1{\texttt{#1}}\fi
\expandafter\ifx\csname urlprefix\endcsname\relax\def\urlprefix{URL }\fi
\expandafter\ifx\csname href\endcsname\relax
  \def\href#1#2{#2} \def\path#1{#1}\fi

\bibitem{Barnesycol1993}
H.~A. Barnes, J.~F. Hutton, K.~Walters, An introduction to Rheology, Rheology
  Series, vol. 3, Ed. Elsevier Science Publishers B.V., Holanda, 1993, 1993.

\bibitem{GalindoRosales2018}
F.~J. Galindo-Rosales, Complex Fluids and Rheometry in Microfluidics, Springer
  International Publishing, Cham, 2018, pp. 1--23.

\bibitem{Morrison2001}
F.~A. Morrison, Understanding Rheology, Oxford University Press, USA, 2001.

\bibitem{BarnesMaia2010}
H.~A. Barnes, J.~M. Maia, Rheometry, in: C.~Gallegos (Ed.), Rheology (vol. II),
  Encyclopaedia of life support systems, Eolss Publishers Co. Ltd. United
  Kingdom, 2010, pp. 331--362.

\bibitem{Ovarlez2012}
G.~Ovarlez, Introduction to the rheometry of complex suspensions, in:
  N.~Roussel (Ed.), Understanding the Rheology of Concrete, Woodhead Publishing
  Series in Civil and Structural Engineering, Woodhead Publishing, 2012, pp. 23
  -- 62.

\bibitem{Macosko1994}
C.~W. Macosko, Rheology: Principles, measurements, and applications, Wiley-VCH,
  Inc., United States, 1994.

\bibitem{Birdetal1987a}
R.~B. Bird, R.~C. Armstrong, O.~Hassager, Dynamics of polymer liquids. Volume 1
  - Fluid mechanics. Second edition., John Wiley and Sons Inc., United States,
  1987.

\bibitem{Galindoetal2013}
F.~J. Galindo-Rosales, M.~A. Alves, M.~S.~N. Oliveira, Microdevices for
  extensional rheometry of slow viscosity elastic liquids: a review,
  Microfluidics Nanofluidics 14 (2013) 1--19.

\bibitem{Dealy1998}
J.~Dealy, A.~Giacomin, Sliding plate and sliding cylinder rheometers, Springer
  Netherlands, 1998, pp. 237--259.

\bibitem{Mezger2002}
M.~T. G., The Rheology Handbook: for user of rotational and oscillatory
  rheometers, Ed. Vincentz Verlag, Alemania, 2002, 2002.

\bibitem{Haake_libro}
G.~Schramm, A practical approach to Rheology and Rheometry, Haake GmbH,
  Karlsruhe, Alemania, 2000, 2000.

\bibitem{GalindoRosales2015326}
F.~J. Galindo-Rosales, S.~Mart\'inez-Aranda, L.~Campo-Dea\~no,
  Cork{STF}$\mu$fluidics - a novel concept for the development of eco-friendly
  light-weight energy absorbing composites, Materials \& Design 82 (2015) 326
  -- 334.

\bibitem{FJGRLCDPatent}
F.~J. Galindo-Rosales, L.~Campo-Dea\~no, Composite layer material for dampening
  external load, obtaining process, and uses thereof, wO Patent App.
  PCT/IB2015/057399 (Sep.~25 2015).

\bibitem{galindoApplSci2016}
F.~J. Galindo-Rosales, Complex fluids in energy dissipating systems, Applied
  Sciences 6(8) (2016) :206.

\bibitem{Goede2019}
T.~de~Goede, K.~de~Bruin, D.~Bonn, High-velocity impact of solid objects on
  non-newtonian fluids, Scientific Reports 9(1) (2019) :1250.

\bibitem{GURGEN2016312}
S.~Gurgen, W.~Li, M.~C. Kushan, The rheology of shear thickening fluids with
  various ceramic particle additives, Materials \& Design 104 (2016) 312 --
  319.

\bibitem{GURGEN2018}
S.~Gurgen, An investigation on composite laminates including shear thickening
  fluid under stab condition, Journal of Composite Materials 53~(8) (2019)
  1111--1122.

\bibitem{KHODADADI2019643}
A.~Khodadadi, G.~Liaghat, S.~Vahid, A.~Sabet, H.~Hadavinia, Ballistic
  performance of kevlar fabric impregnated with nanosilica/peg shear thickening
  fluid, Composites Part B: Engineering 162 (2019) 643 -- 652.

\bibitem{Pinto2017}
F.~Pinto, M.~Meo, Design and manufacturing of a novel shear thickening fluid
  composite ({STFC}) with enhanced out-of-plane properties and damage
  suppression, Applied Composite Materials 24~(3) (2017) 643--660.

\bibitem{Ewoldt2015}
R.~Ewoldt, M.~Johnston, L.~Caretta, Experimental Challenges of Shear Rheology:
  How to Avoid Bad Data, Springer New York, New York, NY, 2015, pp. 207--241.

\bibitem{nomenclature}
Official symbols and nomenclature of the society of rheology, Journal of
  Rheology 57~(4) (2013) 1047--1055.

\bibitem{Giacomin1989}
A.~J. Giacomin, T.~Samurkas, J.~M. Dealy, A novel sliding plate rheometer for
  molten plastics, Polymer Engineering \& Science 29~(8)  499--504.

\bibitem{ORTMAN2011884}
K.~C. Ortman, N.~Agarwal, A.~P. Eberle, D.~G. Baird, P.~Wapperom, A.~J.
  Giacomin, Transient shear flow behavior of concentrated long glass fiber
  suspensions in a sliding plate rheometer, Journal of Non-Newtonian Fluid
  Mechanics 166~(16) (2011) 884 -- 895, papers Presented at the University of
  Wales Institute of Non-Newtonian Fluid Mechanics Meeting on Rheometry, 29-31
  March 2010, Lake Vyrnwy, Wales.

\bibitem{CLASEN20041}
C.~Clasen, G.~H. McKinley, Gap-dependent microrheometry of complex liquids,
  Journal of Non-Newtonian Fluid Mechanics 124~(1) (2004) 1 -- 10.

\bibitem{Moon2008}
D.~Moon, A.~J. Bur, K.~B. Migler, Multi-sample micro-slit rheometry, Journal of
  Rheology 52~(5) (2008) 1131--1142.

\bibitem{Verbaan2015}
C.~A.~M. Verbaan, G.~W.~M. Peters, M.~Steinbuch, Linear viscoelastic fluid
  characterization of ultra-high-viscosity fluids for high-frequency damper
  design, Rheologica Acta 54~(8) (2015) 667--677.

\bibitem{Hatzikiriakos1991}
S.~G. Hatzikiriakos, J.~M. Dealy, Wall slip of molten high density
  polyethylene. i. sliding plate rheometer studies, Journal of Rheology 35~(4)
  (1991) 497--523.

\bibitem{debruyn}
R.~A. Secco, M.~Kostic, J.~R. deBruyn, Fluid Viscosity Measurement, Imprint CRC
  Press, 2018, pp. 1--31.

\bibitem{BIKERMAN194875}
J.~Bikerman, A penetroviscometer for very viscous liquids, Journal of Colloid
  Science 3~(2) (1948) 75 -- 85.

\bibitem{HirtN1981}
C.~Hirt, B.~Nichols, Volume of fluid (vof) method for the dynamics of free
  boundaries, Journal of Computational Physics 39~(1) (1981) 201--225.

\bibitem{Andersson2010}
P.~Andersson, Tutorial multiphaseinterfoam for the dambreak4phase case, Tech.
  rep., Chalmers University of Technology, solid and Fluid Mechanics (2010).

\bibitem{Weller1998}
H.~G. Weller, G.~Tabor, H.~Jasak, C.~Fureby, A tensorial approach to
  computational continuum mechanics using object-oriented techniques, J.
  Computers in Physics 12~(6) (1998) Nov/Dec.

\bibitem{Rafael2017}
R.~Soares~Duarte, Mechanical characterization of shear thickening fluids,
  Master's thesis, University of Minho. School of Engineering (2017).

\end{thebibliography}

\end{document}